\documentclass[12pt]{article}

\usepackage[textheight=22.5truecm,textwidth=16.8truecm,top=2.95truecm,left=2.15truecm]{geometry}
\usepackage{
amsmath,
amsthm,
amssymb,
ascmac,
bm,
tikz,
braket,
mathrsfs,
graphicx,
color,
hyperref,
cite,
titlesec,
caption,
subcaption,
enumerate,
physics,
comment,
array,
empheq,
}
\usepackage[mathscr]{euscript}

\hypersetup{colorlinks,bookmarksopen,bookmarksnumbered,citecolor=black,linkcolor=black,linktocpage,pdfstartview=FitH,urlcolor=black}

\numberwithin{equation}{section}

{\makeatletter
\long\def\@makefntext#1{\parindent 1em\noindent 
\@hangfrom{\hbox to 1.8em{\hss$^{\@thefnmark}$}}#1}
\makeatother}

\def\fnum@figure{\textbf{\figurename\nobreakspace\thefigure}}
\def\fnum@table{\textbf{\tablename\nobreakspace\thetable}}
\long\def\@makecaption#1#2{%
  \vskip\abovecaptionskip
  \sbox\@tempboxa{\small #1. #2}%
  \ifdim \wd\@tempboxa >\hsize
    \small #1. #2\par
  \else
    \global \@minipagefalse
    \hb@xt@\hsize{\hfil\box\@tempboxa\hfil}%
  \fi
  \vskip\belowcaptionskip}
\renewcommand{\l}[0]{\left}
\renewcommand{\r}[0]{\right}

\newcommand{\ie}{{\it i.e.\hspace{0mm}}}
\newcommand{\sdist}[3]{\Omega_{#1}\qty(\bm #2,\bm #3)}
\newcommand{\hdist}[3]{H_{#1}\qty(\bm #2,\bm #3)}
\newcommand{\coinFig}[1]{\includegraphics[width = 0.3\columnwidth]{coin_#1.pdf}}
\newcommand{\disitem}[1]{\vspace{6pt}\noindent {\bf #1}}

\allowdisplaybreaks

\title{\hfill\parbox{3cm}{\normalsize KUNS-2948}\\[12pt]
Bulk reconstruction of AdS$_{d+1}$ metrics\\ and developing kinematic space
}
\author{
Kakeru Sugiura\footnote{sugiura@gauge.scphys.kyoto-u.ac.jp}\qquad\quad
Daichi Takeda\footnote{takedai@gauge.scphys.kyoto-u.ac.jp}\\[12pt]
 \textit{Department of Physics, Kyoto University, Kyoto 606-8502, Japan}
}
\date{}

\begin{document}
\maketitle
\begin{abstract}
    The metrics of the global, Poincar\'e, and Rindler AdS$_{d+1}$ are explicitly reconstructed with given lightcone cuts.
    We first compute the metric up to a conformal factor with the lightcone cuts method introduced by Engelhardt and Horowitz.
    While a general prescription to determine the conformal factor is not known, we recover the factor by identifying the causal information surfaces from the lightcone cuts and finding that they are minimal.
    In addition, we propose a new type of kinematic space as the space of minimal surfaces in AdS$_{d+1}$, where a metric is introduced as a generalization of the case of $d=2$.
    This metric defines the set of bulk points, which is equivalent to that of lightcone cuts.
    Some other properties are also studied towards establishing a reconstruction procedure for general bulk metrics.
\end{abstract}

\newpage

\setcounter{tocdepth}{2}
\tableofcontents
\section{Introduction}\label{sec: introduction}
In the AdS/CFT correspondence, the bulk reconstruction of metrics is a program to understand the bulk spacetime and its dynamics, in terms of the boundary QFT.
In this sense, the completion of the program will help us reveal the origin of gravity, and even quantum theory of it.
Among those, the method with the lightcone cuts (cuts for short)\footnote{
The future (past) lightcone cut of a bulk point $p$ is the intersection between its future (past) lightcone and the conformal boundary (see Fig.\,\ref{fig: the cut} in p.\pageref{fig: the cut}).
} by Engelhardt and Horowitz \cite{Engelhardt:2016wgb,Engelhardt:2016crc} is notable, since it defines the bulk region causally connected to the conformal boundary, and for the region, reconstructs the conformal metric (the metric up to a non-rigid conformal factor).
In addition, their method is covariantly formulated and independent of the dimension.
The extension and application of the method are seen in \cite{Engelhardt:2016vdk,Hernandez-Cuenca:2020ppu,Burda:2018rpb,Folkestad:2021kyz,Takeda:2021dsl}.

In contrast to those achievements, there are still mainly two subjects to be addressed, in order to accomplish a complete reconstruction procedure with the cuts.
First, there is no general prescription to determine the conformal factor of the metric, in spite of several approaches \cite{Engelhardt:2016wgb,Engelhardt:2016crc,Folkestad:2021kyz,Takeda:2021dsl}.
The set of cuts alone do not have any information about the conformal factor, as they are objects related to the causal structure of the bulk.
Second, to obtain the set of cuts from the boundary theory by following their idea of using the bulk-point singularity \cite{Maldacena:2015iua}, we have to repeatedly compute $(d+3)$-point boundary correlators ($d$ is the dimension of the boundary).
Though the proposal guarantees that cuts are in principle computed from the boundary, it is computationally expensive and also has some restrictions.

In this paper, focusing on the pure AdS$_{d+1}$, for which analytic calculations are possible for any $d$, we approach those two subjects; 1) the identification of the conformal factor, and 2) another possibility to find cuts from the boundary.

\begin{figure}
    \centering
    \includegraphics[height=4cm]{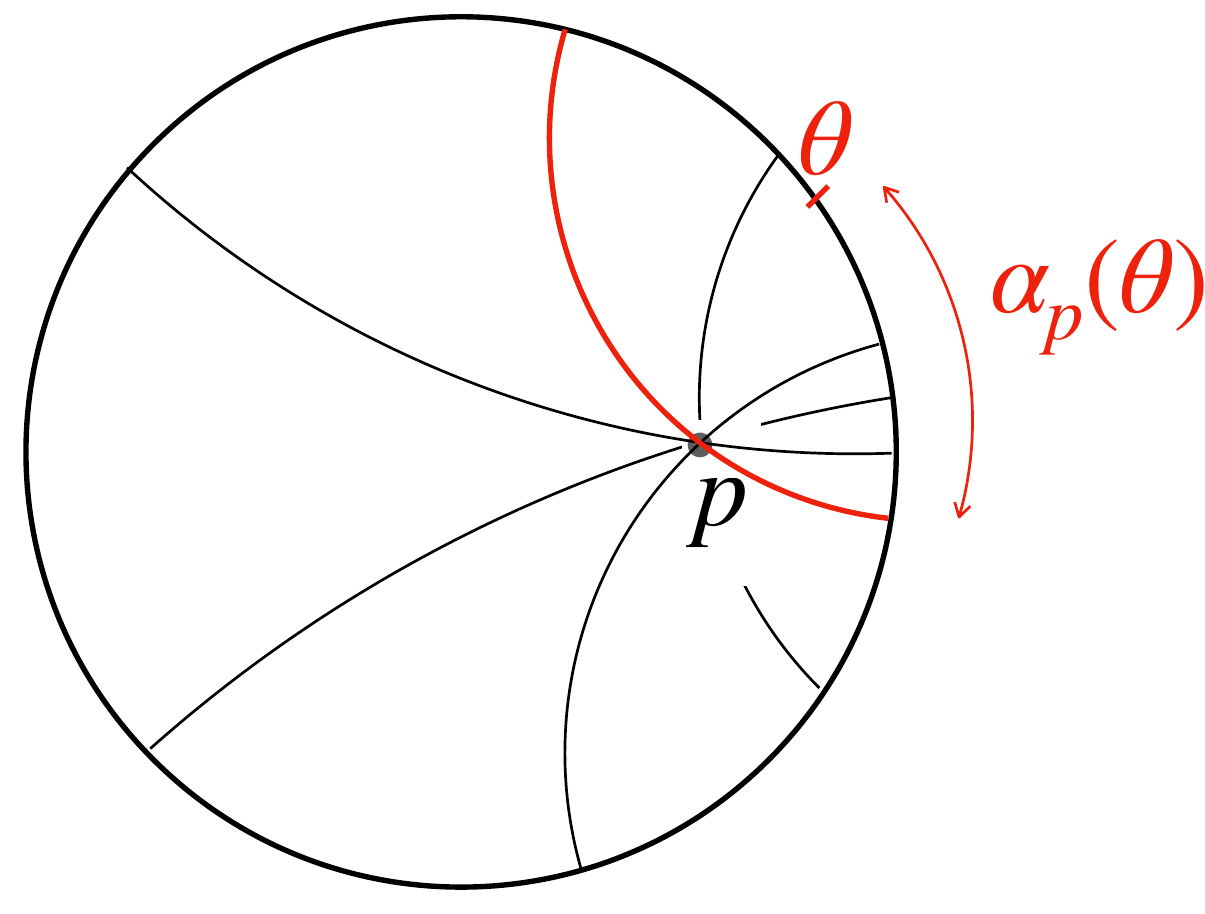}
    \caption{The family of geodesics passing through $p$ on a time slice.
    Each geodesic is labeled by $(\theta,\alpha_p(\theta))$.}
    \label{fig: hole-ography}
\end{figure}

Those questions have partially been solved in AdS$_3$/CFT$_2$.
In \cite{Takeda:2021dsl}, when the causal information surface (CIS)\footnote{
The bifurcation surface of the causal wedge for a boundary region \cite{Hubeny:2012wa}.
} and the minimal surface (a geodesic in 3-dimension) for a boundary interval on a time slice coincide, it was shown that the hole-ography \cite{Balasubramanian:2013lsa} enables us to obtain cuts and determine the missing conformal factor easily.
The hole-ography provides a dictionary which measures bulk curves from the boundary entanglement \cite{Balasubramanian:2013lsa,Balasubramanian:2014sra,Headrick:2014eia}, and a holographic definition of bulk points and distances \cite{Czech:2014ppa}.
The holographic definition of points regards a bulk point $p$ on a time slice as a family of inextensible geodesics which pass through $p$ (Fig.\,\ref{fig: hole-ography}).
Labeling each geodesic by its two endpoints $\theta\pm\alpha$, we see that $\alpha$ is given as a function of $\theta$ such that the corresponding geodesic passes through $p$ ($\alpha_p(\theta)$ in Fig.\,\ref{fig: hole-ography}).
Ref.\cite{Czech:2014ppa} revealed that $\alpha_p$ satisfies an ODE, which can be rewritten by the entanglement entropy of the boundary.
Thus, starting from solving the ODE, we can regard the set of solutions as the time slice of the bulk.
In \cite{Takeda:2021dsl}, the set of cuts was found through the ODE and the condition of CISs being minimal.
After reconstructing the conformal metric from the cuts, the conformal factor is determined by the holographic definition of distances provided by the hole-ography.

In this paper, we will generalize the procedure of \cite{Takeda:2021dsl} to higher dimensional cases.\footnote{
A similar philosophy to ours has already been proposed in \cite{Roy:2018ehv}, where a scalar field is first reconstructed through the modular Hamiltonian without the knowledge of the bulk geometry \cite{Kabat:2017mun}, and the conformal metric is determined from two-point functions.
In the procedure to obtain the scalar field, we get minimal surfaces directly, and the conformal factor can be recovered with the Ryu-Takayanagi formula as we will do in this paper.
In \cite{Roy:2018ehv}, the method was explicitly performed for $d=2$. The application to the higher dimensions is straightforward but technically difficult.
}
We will reconstruct, as examples, the global, Poincar\'e, and Rindler AdS$_{d+1}$ with given lightcone cuts, recovering the missing conformal factor from the Ryu-Takayanagi formula \cite{Ryu:2006bv,Ryu:2006ef}.
So far, the computation of the conformal metric of AdS$_{d+1}$ from the given cuts has not been done explicitly, thus we will first do it here.
After that, the CIS for each ball-shaped region on the boundary will be obtained from the cuts.
Then, we can determine the left conformal factor with the Ryu-Takayanagi formula, because we can find that those CISs are minimal.

Regarding that the CISs are minimal, several holographic interpretations proposed so far are available.
The relation between the CIS and minimal surface for a boundary region was studied in \cite{Hubeny:2012wa,Wall:2012uf,Headrick:2014cta,Cardy:2016fqc}, and on the other hand, the one-point entropy \cite{Kelly:2013aja} is considered to be the dual observable to the area of the CIS (the causal holographic information \cite{Hubeny:2012wa}).
With those tools, we will mention possible ways to holographically show that CISs for balls are minimal.
In the reconstruction of the global AdS$_{d+1}$, there is an easier way to show it and we will demonstrate it.

The second purpose of this paper is to develop another holographic derivation of the cuts like the case of AdS$_3$/CFT$_2$ explained above.
In that case, the hole-ography helped us get the cuts, but now in higher dimensions, the similar differential equation to define the bulk points has not yet been found.
We propose a candidate for it with a new type of kinematic space.

The kinematic space is usually defined as the space of inextensible geodesics in mathematics (each point on the kinematic space corresponds to a geodesic).
In particular, the kinematic space for AdS$_3$/CFT$_2$ \cite{Czech:2015qta,Czech:2015kbp,Czech:2019hdd}, which we write as $K_2$, was introduced as the space of boundary-to-boundary geodesics on a bulk time slice (thus the dimension of the space is two), and a metric is also defined on $K_2$.
Thanks to that, the holographic definition of distances in the hole-ography, was reformulated in terms of volumes on $K_2$.
Moreover, the differential equation for $\alpha_p$'s can be understood as the geodesic equation on $K_2$.
Thus, geodesics on $K_2$ are called ``point-curves" and the set of them is a holographic definition of the bulk points.

In this paper, we will consider in AdS$_{d+1}$ the set of minimal surfaces for boundary ball-shaped regions as a new kinematic space $K_d$.
Since $K_{d}$ can be labeled by the ball radius and center, the dimension of $K_{d}$ is $d$.
We will also define a metric there as a simple generalization of that of $K_2$.
Collecting balls whose minimal surfaces in the bulk pass through a point $p$, we can draw a $(d-1)$-dimensional region on $K_d$ --- we call this region ``point-surface".
We will show that point-surfaces are extremal in $K_d$, thus the Euler-Lagrange equation (the extremal condition) will provide a holographic definition of the bulk points in higher dimensions (to accomplish it, however, we have to write the metric on $K_d$ in terms of the boundary theory, as has already been done for $d=2$).

The organization of this paper is as follows.
In \S\ref{sec: preparation}, we prepare for our reconstruction strategy, including review. 
In addition, we show how a CIS can be identified from the set of cuts.
In \S\ref{sec: reconstruction}, the reconstruction procedure is introduced, and we apply it to explicitly reconstruct AdS$_{d+1}$ in three different patches.
In \S\ref{sec: kinematic space}, we introduce our kinematic space and study its properties.
We devote \S\ref{sec: discussion} to summary and discussions.
In appendix \ref{app: calculation detail}, the computation in the reconstruction of \S\ref{sec: reconstruction} is shown more in detail.
In appendix \ref{app: Sch-AdS}, we numerically check $\mathcal{C}_A = \mathcal{E}_A$ in Schwarzschild-AdS$_{d+1}$.

\section{Preparation for reconstruction}\label{sec: preparation}
In this section, we prepare for our reconstruction scenario, including review of previous works.
In \S\ref{subsec: cuts}, we review the lightcone cuts method \cite{Engelhardt:2016wgb}, discuss the things to overcome, and then compute the cuts in the pure AdS$_{d+1}$ from bulk analyses.
Then in \S\ref{subsec: surfaces}, we review the definition of the causal information surface (CIS) and its relation with the minimal surface.
Finally in \S\ref{subsec CIS from cuts}, we show that the CIS for a boundary region can be obtained from the set of cuts.
This will be used in \S\ref{sec: reconstruction}.

\subsection{Lightcone cuts}\label{subsec: cuts}
\subsubsection{Review of the lightcone cuts method}\label{subsubsec: review of cuts}

\begin{figure}
    \centering
    \includegraphics[height = 5cm]{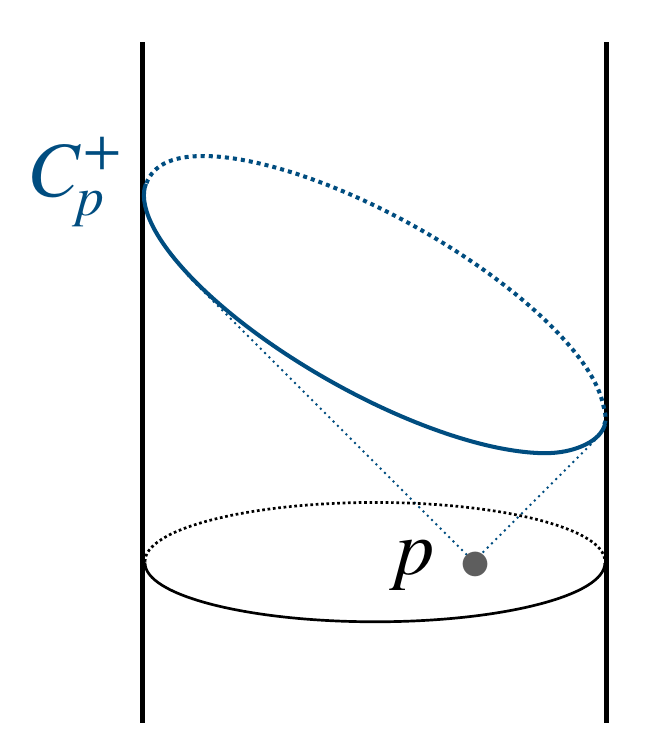}
    \caption{The future lightcone cut of $p$.}
    \label{fig: the cut}
\end{figure}

Let $M$ be an asymptotically AdS spacetime\footnote{More technical assumptions are imposed on $M$ to rigorously show the facts here (see \cite{Engelhardt:2016wgb}).} of $d+1$ dimensions, and $J^\pm(p)$ be the causal future/past of $p\in M$.
The future/past lightcone cut of $p$, $C_p^\pm$, is defined as
\begin{align}
    C_p^\pm := \partial J^\pm(p)\cap \partial M,
\end{align}
with $\partial$ meaning the boundary (especially, $\partial M$ denotes the conformal boundary).
The future cut is depicted in Fig.\,\ref{fig: the cut}.
Under this definition, the following holds ($I^\pm$ denotes chronological future/past):
\begin{enumerate}
    \item For $p\in I^-(\partial M)\,(I^+(\partial M))$, there exists a unique future (past) cut of $p$.\label{property: cuts existence}
    \item If and only if $p=q$, both $C^+_p\cap C^+_q$ and $C^-_p\cap C^-_q$ are open and non-empty.\label{property: injection}
    \item If $C_p^+$ is tangent to $C_q^+$ at a unique point, then $p$ and $q$ are null separated (the same thing holds for $C_p^-$ and $C_q^-$).\label{property: null vector}
\end{enumerate}
Property \ref{property: cuts existence} and \ref{property: injection} together roughly mean that there is a one-to-one map between $I^-(\partial M)$ and the set of future cuts, as well as between $I^+(\partial M)$ and the set of past cuts.
Thus, in the bulk reconstruction, the set of cuts can be regarded as the bulk region chronologically connected to $\partial M$.
On the other hand, property \ref{property: null vector} says that the set of cuts knows the bulk null vectors.

It was proposed in \cite{Engelhardt:2016wgb} that the bulk-point singularity \cite{Maldacena:2015iua} can identify cuts.
The bulk-point singularity is a kind of correlator divergences in holographic QFTs.
The divergence happens if there exists a point in the dual bulk such that it is null separated from all points in the correlator and we can assign momenta to propagators while conserving the total momentum.\footnote{
Though the existence of the bulk-point singularity is confirmed only for the pure AdS, its existence and possibility of application to general geometries are also mentioned in \cite{Engelhardt:2016wgb}.
}

By using this, the set of cuts can be obtained in the boundary language as follows.
We first fix $d+1$ points on the boundary so that any two points are spacelike separated.
Next, putting additional boundary points in the chronological future of them, we consider the correlator among those $d+3$ points.
Then, we move the last two points and find their positions where the correlator diverges.
The orbit that the two points draw is the future cut of the holographic interaction point of the correlator.
We repeat this procedure to obtain all future cuts.
With the ``future" and ``past" appearing above exchanged, we do the same thing to obtain the past cuts.

Let us assume that we have constructed the set of future cuts from the boundary.
Because of the property \ref{property: cuts existence} and \ref{property: injection}, we will be able to label the cuts by using $d+1$ continuous parameters $\lambda = (\lambda^0,\cdots,\lambda^d)$ (the same number as the bulk dimension), by which we write the cuts as $C^+(\lambda)$.
Here we choose $\lambda$ so that $C^+(\lambda)$ is smooth with respect to $\lambda$.
At this stage, the set $\{C^+(\lambda)|\lambda\}$ is regarded as a holographic definition of $I^-(\partial M)$ of the bulk and $\lambda$ is interpreted as a coordinate.
The metric to be reconstructed will be written in this coordinate as $g_{\mu\nu}(\lambda)$.

Next, to obtain the metric, we find infinitesimal variations $\delta\lambda = \delta\lambda(\lambda)$ such that $C^+(\lambda)$ and $C^+(\lambda+\delta\lambda)$ are tangent at exactly one point: if we write the cut as $t = C(\lambda,\bm x)$ with $(t,\bm x)$ being the boundary coordinate, the condition is given as
\begin{align}
    C(\lambda,\bm x) = C(\lambda+\delta\lambda,\bm x),\qquad
    \nabla_{\bm x} C(\lambda,\bm x) = \nabla_{\bm x} C(\lambda+\delta\lambda,\bm x).
        \label{eq: tangentcuts}
\end{align}
The property \ref{property: null vector} claims that all of such infinitesimal $\delta\lambda$ be a null vector at $\lambda$.
Thus, the metric must satisfy
\begin{align}
    \delta\lambda^\mu \delta\lambda^\nu g_{\mu\nu}(\lambda) = 0
    \qquad
    (\mu,\nu = 0,\cdots,d).\label{eq: homogeneous equation}
\end{align}
The metric can be determined through the above equation by collecting enough number of such $\delta\lambda$'s for each $\lambda$.
However, the fact that \eqref{eq: homogeneous equation} is homogeneous makes it impossible to fix the normalization of $g_{\mu\nu}(\lambda)$, and hence all we can do is to collect $d(d+3)/2$ null vectors and solve for $g_{\mu\nu}(\lambda)$ up to a conformal factor which may depend on $\lambda$.
If we would also like to cover $I^+(\partial M)$, we do the same thing for the past cuts.
There are several proposals to determine the conformal factor after we have reconstructed the conformal metric, but no general method which is valid for any geometries has not been found.
In this paper, we study another possibility and explicitly reconstruct the pure AdS$_{d+1}$.

\subsubsection{On the holographic derivation of lightcone cuts}\label{subsubsec: derivation of cuts}
On deriving cuts by using the bulk-point singularity, there are several points to be addressed.
\begin{itemize}
    \item Since we use boundary-to-boundary correlators, all cuts we can get are those of bulk points in $I^+(\partial M)\cap I^-(\partial M)$ (points which can connect to the boundary in both the chronological future and past).
    Thus, for example, we cannot obtain any cut of the BTZ black hole, because all null geodesics shot from the boundary go down to the horizon, never to come back again.
    \item The existence of the bulk-point singularity is assured only in the global patch. For example, in pure Poincar\'{e} AdS, any null geodesic shot from the boundary has a momentum whose radial part directs away from the boundary, so the momentum conservation condition necessary for the singularity to happen cannot be satisfied.
    \item Carrying out the idea of using the bulk-point singularity is so tough. Even in CFT$_2$, it is not easy to compute 5-point functions.
\end{itemize}

In \cite{Takeda:2021dsl}, it was shown that we are able to obtain the cuts easily through the hole-ography (or the kinematic space) in holographic QFTs dual to locally AdS$_3$ geometries.
Thus, the combination of the cuts and hole-ography seems effective, but there is still no similar thing known in higher-dimensions, which is our motivation to develop a new kinematic space in \S\ref{sec: kinematic space}.
We will also see that the story of the kinematic space seems independent of the choice of the patch.

\subsubsection{Computation of the cuts in the bulk}\label{subsubsec: cuts from bulk}
Here from the viewpoint of the bulk, we compute the lightcone cuts for the pure AdS$_{d+1}$, for the Poincar\'e, global, and Rindler patches.

\subsubsection*{Poincar\'e patch}
The geometry is given as
\begin{align}
    ds^2 = \frac{L^2}{z^2}(-dt^2 + d\bm x^2 + dz^2),\qquad \bm x = (x^1,\cdots,x^{d-1}).
\end{align}
Since the metric is conformally flat, the lightcones are identical to that of the Minkowski spacetime.
Let $(t,\bm x,z) = (T,\bm X,Z)$ be a bulk point.
Its lightcone is written as
\begin{align}
    (t-T)^2 - (\bm x-\bm X)^2 - (z-Z)^2 = 0.
\end{align}
Taking $z\to +0$, we obtain the future/past cut as
\begin{align}
    t = T\pm \sqrt{(\bm x-\bm X)^2 + Z^2}.\label{eq: Poincare cuts}
\end{align}

\subsubsection*{Global patch}
The metric is given as
\begin{align}
    ds^2 &= -\frac{r^2+L^2}{L^2}dt^2 + r^2d\Omega_{d-1}^2 + \frac{L^2dr^2}{r^2+L^2},\quad
    d\Omega_{d-1}^2 = \sum_{k=1}^{d-1}\sin^2\theta^1\cdots\sin^2\theta^{k-1}(d\theta^k)^2,
\end{align}
where $d \Omega_{d-1}$ is the line element on $\mathbb{S}^{d-1}$.
Let $(t,\bm \theta,r) = (T,\bm \Theta,R)$ be a bulk point (we use $\bm \theta$ to collectively handle $\theta^k$'s).
A future/past null geodesic starting from this point along the $\theta^1$-direction, is determined by solving the following equations:
\begin{align}
    \dot t = \pm 1,\quad
    r^2\dot \theta^1 = \text{const.},\quad
    \frac{r^2+L^2}{L^2}\dot t^2= + r^2(\dot\theta^1)^2 + \frac{L^2\dot r^2}{r^2+L^2},\quad
    (t,\bm\theta,r)|_{\lambda=0} = (T,\bm\Theta,R).
\end{align}
Here $\lambda$ is an affine parameter, and the dot means the $\lambda$-derivative.
The equations are easy to solve because they are the same as those of the AdS$_3$.
After removing the constant appearing above (the angular momentum), we obtain the one-parameter family of endpoints ($\theta^1$ parametrizes them):
\begin{align}
    t = T\pm L\cos^{-1}\l(\frac{R}{\sqrt{R^2+L^2}}\cos(\theta^1-\Theta^1)\r).\label{eq: 3D cuts in global}
\end{align}

To obtain the whole lightcone cuts, we must take all null geodesics into account.
However, it is always possible to align any initial velocity in the $\theta^1$-direction by rotating the sphere with the initial point fixed.
Then we obtain the same expression as \eqref{eq: 3D cuts in global}, but in the new coordinate.
Going back to the original coordinate, we see that $|\theta^1-\Theta^1|$ appearing in \eqref{eq: 3D cuts in global} is replaced with the geodesic distance between $\bm\theta$ and $\bm \Theta$ along $\mathbb{S}^{d-1}$, which we write as $\sdist{d-1}{\theta}{\Theta}$ hereafter.
Thus, the whole future/past cut of $(T,\bm \Theta,R)$ is given as
\begin{align}
    t=T\pm L\cos^{-1}\qty(\frac{R}{\sqrt{R^2+L^2}}\cos\sdist{d-1}{\theta}{\Theta}),\label{eq: Global cuts}
\end{align}

In general, to express the geodesic distance between $(\theta^{k},\cdots,\theta^{d-1})$ and $(\Theta^{k},\cdots,\Theta^{d-1})$ on the unit $\mathbb{S}^{d-k}$, we use $\sdist{d-k}{\theta}{\Theta}$.
Then, we have a recurrence formula,
\begin{align}
    \cos\sdist{d-k+1}{\theta}{\Theta} = \cos\theta^k\cos\Theta^k+\sin\theta^k\sin\Theta^k\cos\sdist{d-k}{\theta}{\Theta}.\label{eq: sdist relation}
\end{align}
which can straightforwardly be shown by considering $\mathrm{SO}(d)$ rotation of $\mathbb{S}^{d-1}$.

\subsubsection*{Rindler patch}
The metric is given as
\begin{align}
    ds^2 &= -\frac{r^2-L^2}{L^2}dt^2 + r^2dH_{d-1}^2 + \frac{L^2dr^2}{r^2-L^2},\quad
    d H_{d-1}^2 = \sum_{k=1}^{d-1}\sinh^2\chi^1\cdots\sin^2\chi^{k-1}(d\chi^k)^2,
\end{align}
where $d H_{d-1}$ is the line element on the hyperbolic space $\mathbb{H}^{d-1}$.
Let $(t,\bm \chi,r) = (T,\bm X,R)$ be a bulk point (we use $\bm \chi$ to collectively handle $\chi^k$'s).
The lightcone cuts of the future/past cut of $(T,\bm X,R)$ is obtained in the same way as before:
\begin{align}
    t=T\pm L\cosh^{-1}\qty(\frac{R}{\sqrt{R^2-L^2}}\cosh\hdist{d-1}{\chi}{X}).\label{eq: Rindler cuts}
\end{align}
Here $\hdist{d-1}{\chi}{X}$ denotes the geodesic distance on $\mathbb H^{d-1}$, and is given by
\begin{align}
    \cosh\hdist{d-1}{\chi}{X} = \cosh\chi^1\cosh X^1 - \sinh\chi^1\sinh X^1\cos\sdist{d-2}{\chi}{X}.\label{eq: hdist relation}
\end{align}

\subsection{Causal information surface}\label{subsec: surfaces}

\begin{figure}[t]
    \centering
    \includegraphics[height=6cm]{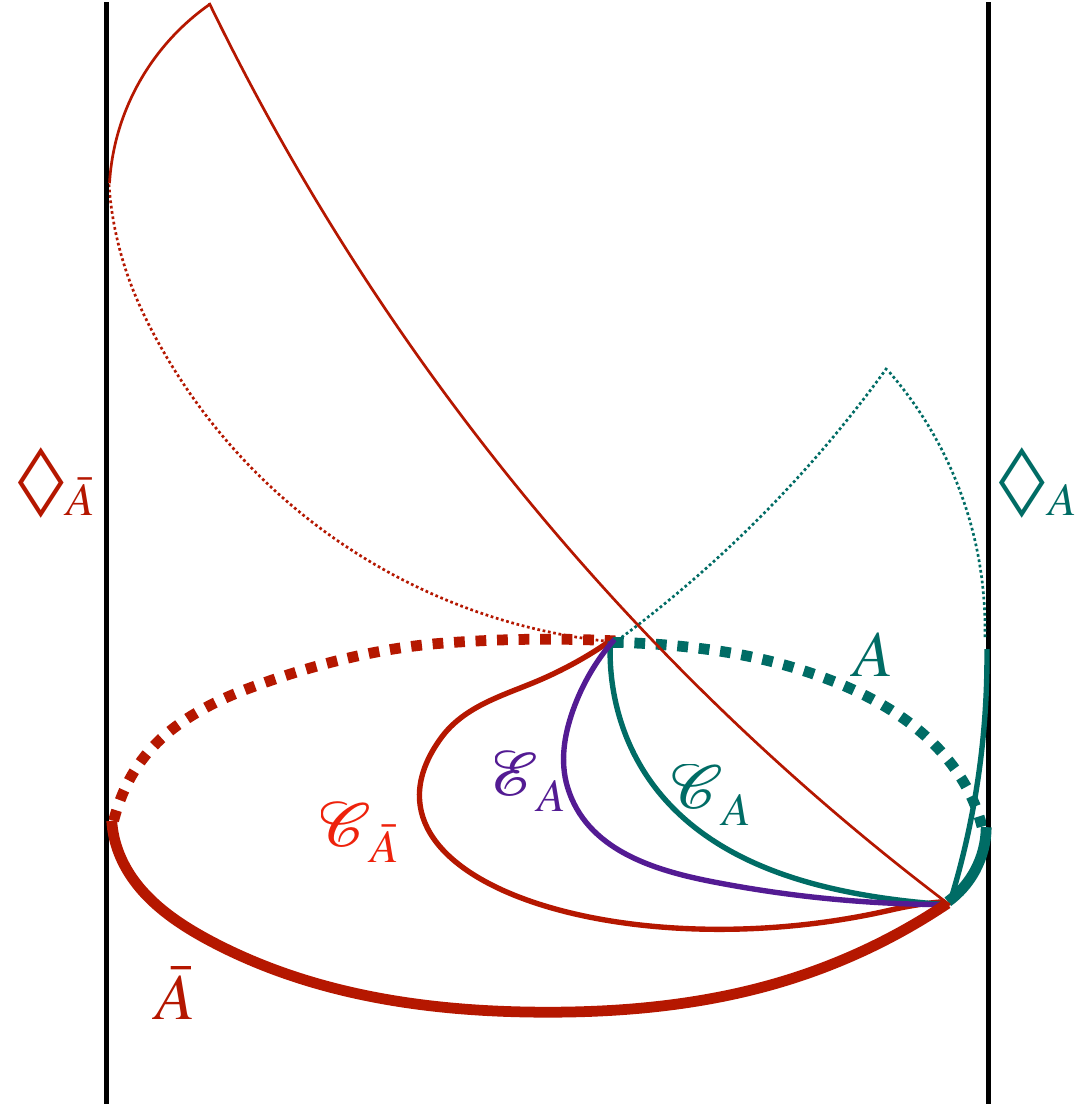}
    \caption{The position of $\mathcal{C}_A$, $\mathcal{C}_{\bar A}$, and $\mathcal{E}_A$.
    The lower halves of $\Diamond_A$ and $\Diamond_{\bar A}$ are omitted in the figure.}
    \label{fig: surfaces}
\end{figure}

Here, we review the definition and properties of the causal information surface (CIS) \cite{Hubeny:2012wa}.
Let $A$ be a boundary spatial region and $\Diamond_A$ be its boundary domain of dependence.
The causal wedge of $A$, $\blacklozenge_A$ is defined as
\begin{align}
    \blacklozenge_A := J^-(\Diamond_A)\cap J^+(\Diamond_A),\label{eq: causal wedge}
\end{align}
and the CIS of $A$, $\mathcal{C}_A$, is defined as
\begin{align}
        \mathcal{C}_A := \partial J^-(\Diamond_A)\cap \partial J^+(\Diamond_A).\label{eq: CIS}
\end{align}

Let $\mathcal{E}_A$ be the minimal surface of $A$ (strictly speaking, the maximin surface \cite{Wall:2012uf}).
    With some conditions including the null curvature condition, $\mathcal{E}_A$ generally lies outside $\blacklozenge_A$ \cite{Hubeny:2012wa, Wall:2012uf, Headrick:2014cta}.
The question as to when $\mathcal{C}_A$ equals $\mathcal{E}_A$ has not yet been answered.
In \cite{Hubeny:2012wa}, $\mathcal{C}_A = \mathcal{E}_A$ ($A:$ any ball) was confirmed, by analyzing the bulk, for some pure Einstein geometries; the Poincar\'e AdS, global AdS, and BTZ.
We will numerically show that the coincidence holds also for higher-dimensional AdS black holes in appendix \ref{app: Sch-AdS}.

The relation between $\mathcal{C}_A$ and $\mathcal{E}_A$ has been studied also from the boundary viewpoint.
The area of $\mathcal{C}_A$ in the unit of $4G$ ($G:$ the Newton constant) is called causal holographic information \cite{Hubeny:2012wa}, which in \cite{Kelly:2013aja} was conjectured to be equivalent to the one-point entropy in the boundary theory.
On the other hand, we have by definition $\mathrm{Area}(\mathcal{E}_A)\leq \mathrm{Area}(\mathcal{C}_A)$, and it was  conjectured in \cite{Hubeny:2012wa} that the saturation of the area, $\mathrm{Area}(\mathcal{E}_A)= \mathrm{Area}(\mathcal{C}_A)$, occurs when $A$ and $\bar A$ are maximally entangling.

\begin{figure}[t]
    \centering
    \includegraphics[height = 5cm]{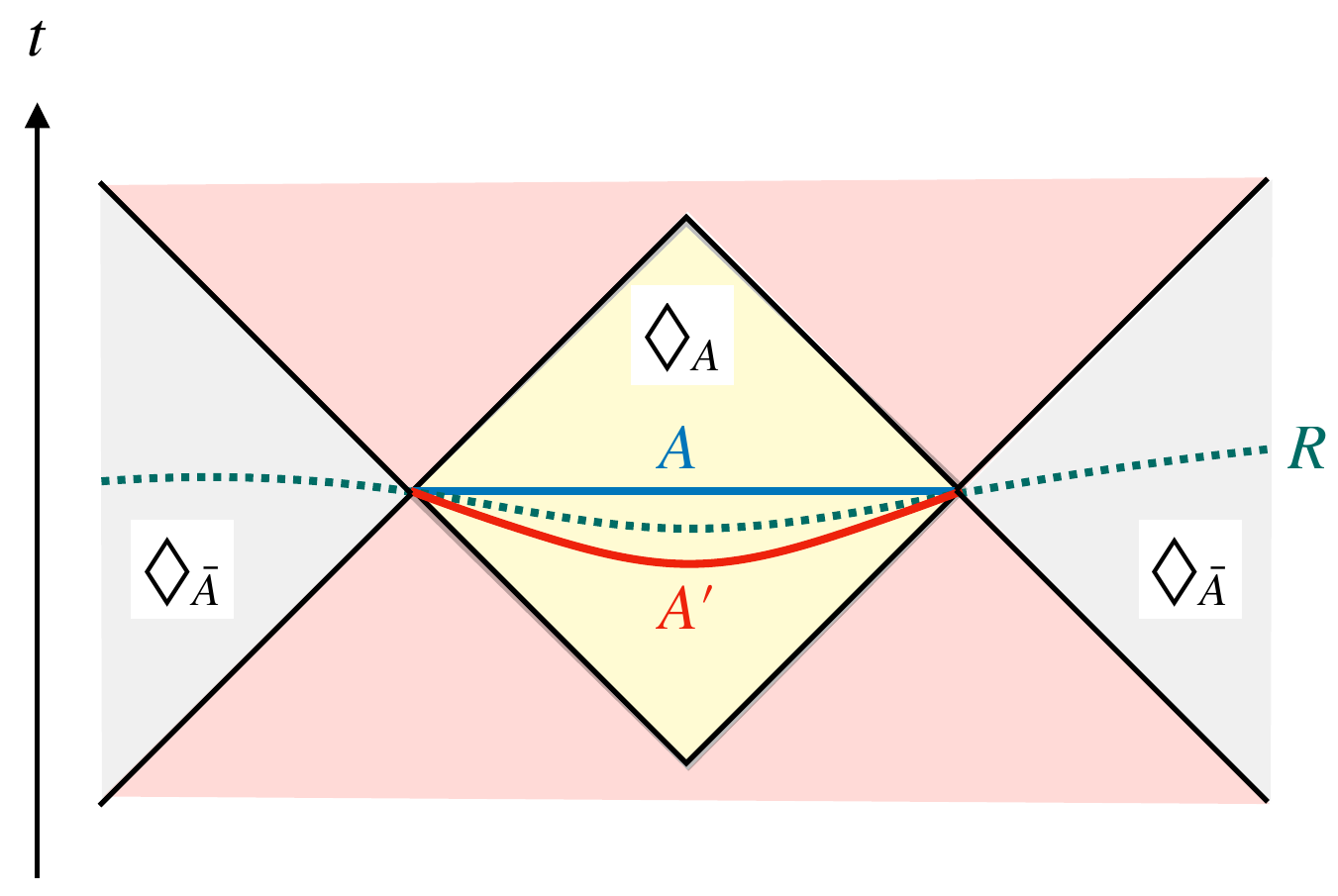}
    \caption{Regions $A$ and $A'$ that satisfy $\Diamond_A = \Diamond_{A'}$. The domain $R$ is the support of the perturbation.}
    \label{fig: diamond}
\end{figure}

On the other hand, the relation about the bulk position among $\mathcal{C}_A$, $\mathcal{C}_{\bar A}$, and $\mathcal{E}_A$ was studied in \cite{Headrick:2014cta} (see Fig.\,\ref{fig: surfaces}).
If the boundary state is pure, $\mathcal{E}_A$ must exist between $\mathcal{C}_A$ and $\mathcal{C}_{\bar A}$ on the bulk Cauchy surface containing them.
This is based on the following consideration.
Let $A'$ be any boundary region such that $\Diamond_A = \Diamond_{A'}$.
Then, the reduced density matrices $\rho_A$ and $\rho_{A'}$ are unitary related \cite{Casini:2003ix}.
Thus, we have $S_A = S_{A'}$ ($S_X$ is the entanglement entropy of region $X$).
This means $\mathcal{E}_A = \mathcal{E}_{A'}$ in the dual bulk, by the HRT formula \cite{Hubeny:2007xt}.
On the other hand, if we first fix the initial state at $t\to-\infty$ and put any perturbation to the Hamiltonian with its support $R\subset\Diamond_A\cup \Diamond_{\bar A}$, we can always find a boundary Cauchy surface including $A'$ such that it is entirely in $J^-(R)$ and $\Diamond_A = \Diamond_{A'}$ (Fig.\,\ref{fig: diamond}).
The perturbation of course does not affect $\rho_{A'}$, thus does not affect $\rho_A$ from the above arguments.
Fixing the final state to consider the time-reversed evolution, we again reach the same conclusion by evaluating $S_{A'}$ on $A'\subset J^+(R)$, instead of $S_A$ itself.
Thus in the bulk, $\mathcal{E}_A$ must not be affected by any perturbation on $\Diamond_A\cup \Diamond_{\bar A}$ from the past or the future.
Therefore, we conclude that the bulk surfaces are located as Fig.\,\ref{fig: surfaces}.

\subsection{Causal information surfaces from lightcone cuts}\label{subsec CIS from cuts}

\begin{figure}[t]
    \centering
    \includegraphics[height = 8cm]{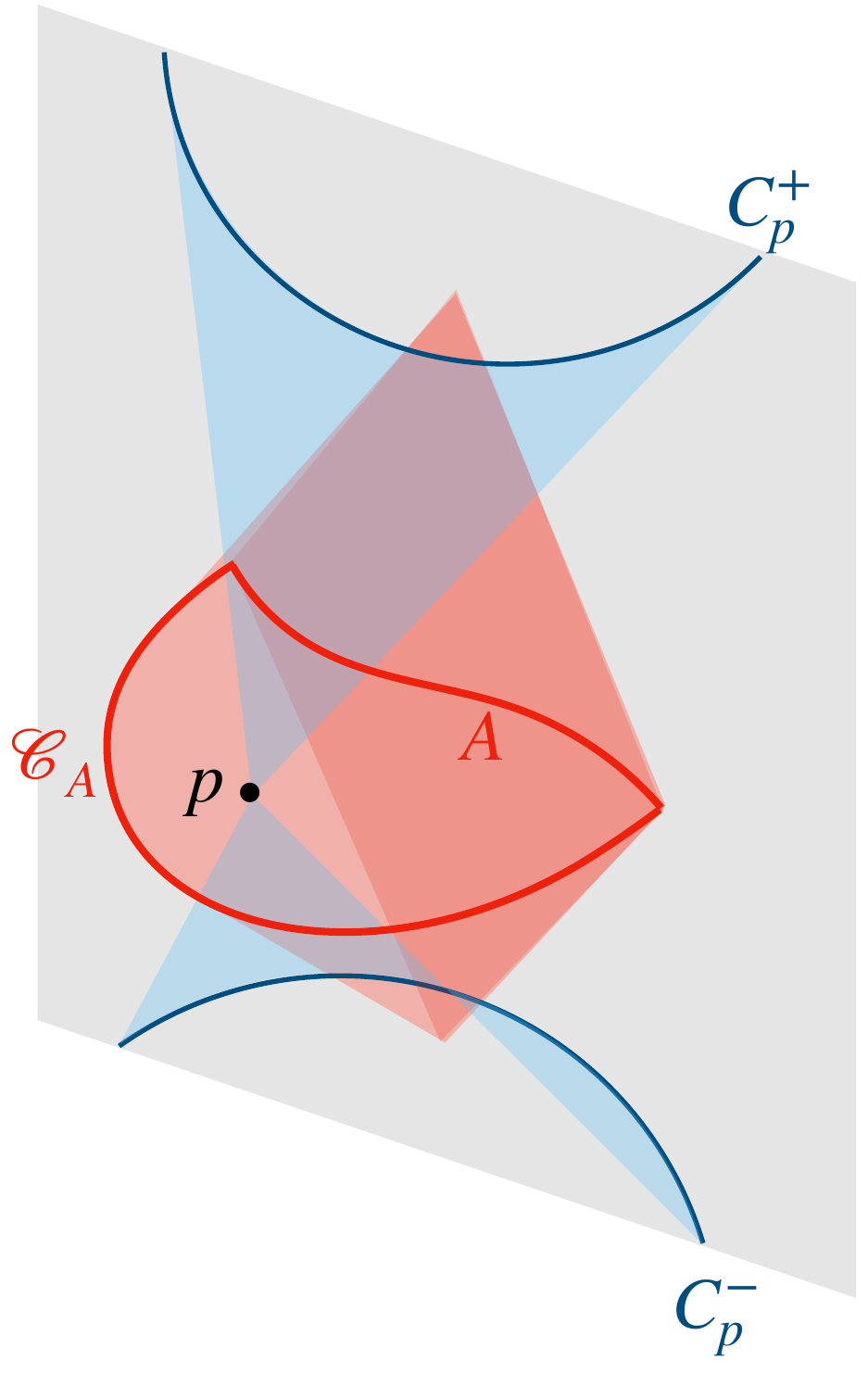}
    \caption{The causal wedge and CIS of region $A$, and how they are obtained from the set of cuts.}
    \label{fig: CIS from cuts}
\end{figure}

Here we show that the CIS for a boundary region can be reconstructed from the set of lightcone cuts.
Let $A$ be a spatial region on the boundary.
For any $p\in \blacklozenge_A$, there exists a null curve connecting $p$ and a point in $\Diamond_A$.
Then, the future lightcone cut of $p$, $C_p^+$, shares at least a point with $\Diamond_A$ (Fig.\,\ref{fig: CIS from cuts}): $C_p^+\cap \Diamond_A\neq \emptyset$.
From the similar argument, we also see $C_p^-\cap \Diamond_A\neq \emptyset$.
The converse argument is also valid.
Thus, we have
\begin{align}
    \blacklozenge_A = \{p|C_p^+\cap \Diamond_A\neq \emptyset \wedge C_p^-\cap \Diamond_A\neq \emptyset\}.
\end{align}
Similarly, by the definition \eqref{eq: CIS}, we also have
\begin{align}
    \mathcal{C}_A = \partial\{p|C_p^+\cap \Diamond_A\neq \emptyset\}\cap \partial\{p|C_p^-\cap \Diamond_A\neq \emptyset\}.
    \label{eq: holographic CIS}
\end{align}
Since the validity of \eqref{eq: holographic CIS} is guaranteed only by the nature of the spacetime manifold, we can use it to find CISs from the boundary theory.\footnote{
After reconstructing the conformal metric by cuts, CISs can be computed from it.
However, the formula or its reduced forms introduced below provide a direct way and work even when it is difficult to reconstruct the conformal metric.
}

In the metric reconstruction in \S\ref{sec: reconstruction}, we only need $\mathcal{C}_A$ for ball-shaped $A$'s.\footnote{
Here balls are defined with respect to the geodesic distance of the boundary.
}
In this case, $\mathcal C_A$ is given as the intersection between the past lightcone of the top vertex of $\Diamond_A$ and the future lightcone of the bottom vertex.
Then, \eqref{eq: holographic CIS} now reads
\begin{align}
    \mathcal{C}_A = \{p|v_A^+\in C_p^+ \wedge v_A^- \in C_p^- \}\qquad
    (\mbox{for spherical $A$}),
    \label{eq: holographic CIS for ball}
\end{align}
where $v_A^\pm$ is the top/bottom vertex of $\Diamond_A$.

To be more in concrete, let us consider a static bulk, take a time slice $t=T$, and write $C_p^\pm$ as $t = T \pm c(\lambda,\bm x)$.
Here $\bm x$ is the symbol for spatial coordinates of the boundary and $\lambda$ is the bulk coordinate of $p$.
If we write the radius of $A$ as $\alpha$ and the center as $\bm x_0$, the coordinate of $v_A^\pm$ is $(T\pm \alpha,\bm x_0)$.
Then, \eqref{eq: holographic CIS for ball} is equivalent to
\begin{align}
    c(\lambda,\bm x_0) = \alpha.
    \label{eq: static CIS}
\end{align}
Seen as a constraint on $\lambda$, this equation is nothing but $\mathcal{C}_A$.

\section{Reconstruction of pure AdS\texorpdfstring{$_{d+1}$}{TEXT} geometries}\label{sec: reconstruction}

In this section, assuming that we have obtained the lightcone cuts from the boundary data, we explicitly reconstruct the metrics of the pure AdS$_{d+1}$ in the global, Poincar\'e, and Rindler patch.
We first explain our reconstruction procedure in \S\ref{subsec: method}, and then apply it to those three geometries.
The subject, how to obtain the lightcone cuts, will be later addressed in \S\ref{sec: kinematic space}.

\subsection{Reconstruction procedure}\label{subsec: method}
Our reconstruction procedure consists of the following three steps.
\vskip\baselineskip
\noindent
{\it \underline{Step 1.} Carry out the lightcone cuts method.}
\vskip.5\baselineskip
\noindent
We follow the lightcone cuts method, which has been reviewed in \S\ref{sec: preparation}, to reconstruct the conformal metric, the metric up to a conformal factor.
The procedure to obtain cuts cannot be performed now, because of the difficulties explained in \S\ref{subsubsec: derivation of cuts}.
We will return to this in \S\ref{sec: kinematic space}.
Here, starting with given cuts, we solve the tangential condition \eqref{eq: tangentcuts} to find null vectors, then reconstruct the conformal metric from \eqref{eq: homogeneous equation}.

\vskip\baselineskip
\noindent
{\it \underline{Step 2.} Build CISs from the lightcone cuts and obtain the minimal surfaces.}
\vskip.5\baselineskip
\noindent
In this step, we build CISs from the set of cuts and find the minimal surfaces from them.
As is explained in \S\ref{sec: preparation}, the CIS for a boundary ball-shaped region $A$, $\mathcal{C}_A$, is built from the set of cuts.
If the system is static, the bulk is also static, and hence we can use \eqref{eq: static CIS}.
The minimal surface $\mathcal E_A$ is supposed to be identified from $\mathcal C_A$, and here, we treat the simple cases where $\mathcal E_A = \mathcal C_A$ for any ball $A$.
This can be shown by the tools explained in \S\ref{subsec: surfaces}.
Especially, if $\mathcal C_A = \mathcal C_{\bar A}$, there is a far simpler prescription to say $\mathcal E_A = \mathcal C_A$; since, due to the boundary causality, the position relation among $\mathcal C_A$, $\mathcal C_{\bar A}$, and $\mathcal E_A$ is generally as depicted in Fig.\,\ref{fig: surfaces}, relation $\mathcal E_A = \mathcal C_A$ immediately follows from $\mathcal C_A = \mathcal C_{\bar A}$.
We will use this strategy in reconstructing the global AdS.

\vskip\baselineskip
\noindent
{\it \underline{Step 3.} Identify the conformal factor with the Ryu-Takayanagi formula.}
\vskip.5\baselineskip
\noindent
The final step is to identify the undetermined conformal factor from the fact that the surface $\mathcal E_A$ obtained in step 2 is minimal.
Let us pick a representative $\tilde g_{\mu\nu}$ from the conformal metric reconstructed in the step 1, and write the exact metric as $g_{\mu\nu}(\lambda)=e^{2\omega(\lambda)}\tilde{g}_{\mu\nu}(\lambda)$, where $\lambda$ is the bulk coordinate.
Since the area of each $\mathcal E_A$ is given in terms of $\omega(\lambda)$, the extremal condition (the Euler-Lagrange equation) provides a differential equation for $\omega(\lambda)$ of $\lambda\in \mathcal E_A$.
For fixed $\lambda$, there are infinite number of $A$'s such that $\lambda\in \mathcal E_A$,\footnote{
There is a region called ``entanglement shadow" \cite{Hubeny:2012ry,Engelhardt:2013tra,Engelhardt:2015dta}, which cannot be probed by minimal surfaces.
However, using non-minimum but extremal surfaces enables us to probe there.
The boundary quantity dual to the area of each of such surface is called ``entwinement" \cite{Balasubramanian:2014sra, Lin:2016fqk, Balasubramanian:2016xho,  Erdmenger:2019lzr, Craps:2022pke}, 
}
and hence $\omega(\lambda)$ can be computed up to overall constant (because the extremal condition is homogeneous).

The remaining constant can be fixed by comparing the area of $\mathcal E_A$ with the dual entanglement entropy $S_A$:
\begin{align}
    S_{A}=\frac{\mathrm{Area}(\mathcal{E}_A)}{4G}.
        \label{eq: RTformula}
\end{align}
Since there is an ambiguity of the choice of the cutoff, it is reasonable to see the cutoff-independent term in the above formula\footnote{
There may be various ways to determine the constant $C$.
For example, if the bulk contains probe fields, $C$ will appear in correlators.
Here we will propose a way closed in the bulk geometry.
} (this is enough because our purpose is just to fix one constant).

If the boundary state is symmetric under some group and we can make an ansatz $\omega = \omega(\lambda^*)$ with $\lambda^*$ being one of the bulk coordinates, then step 3 becomes quite simple as we will see in the following examples.


\subsection{Application to global, Poincar\'e, and Rindler AdS\texorpdfstring{$_{d+1}$}{TEXT}}\label{subsec: explicit reconstruction}
Here, we apply the strategy in \S\ref{subsec: method} to global, Poincar\'e, and Rindler AdS$_{d+1}$ to demonstrate how it works.
We assume that the lightcone cuts have already been constructed.

\subsubsection{Global patch}\label{subsubsec: reconstruct G}
Let us consider the ground state of a holographic CFT defined on $\mathbb R\times\mathbb S^{d-1}$,
\begin{align}
    ds^2_\text{bdy}=-dt^2+L^2 d\Omega_{d-1}^2,
        \label{eq: bdyg in G}
\end{align}
where $L$ is the radius of $\mathbb{S}^{d-1}$.
We suppose that the universal term of the entanglement entropy for any ball $A$ is known to be given as
\begin{subequations}
    \begin{align}
        S_A&= \cdots + \frac{\Omega_{d-2}\,L^{d-1}}{4G}\cdot (-1)^{\frac{d-2}{2}}\frac{(d-3)!!}{(d-2)!!}\ln\qty(\frac{l}{\epsilon}) + \cdots \quad (d\text{: even}),
    \label{eq: even S_A}\\
        S_A&= \cdots + \frac{\Omega_{d-2}\,L^{d-1}}{4G}\cdot \frac{\Gamma\qty(\frac{d-1}{2})\Gamma\qty(\frac{2-d}{2})}{2\sqrt{\pi}} + \cdots \quad (d\text{: odd}).\label{eq: odd S_A}
    \end{align}
    \label{eq: entanglement in G}
\end{subequations}
Here $l$ is some length and $\epsilon$ is a cutoff having the dimension of length.
The cutoff-independent part is the coefficient of the logarithm in even $d$ and is the constant term in odd $d$.

\vskip\baselineskip
\noindent
{\it \underline{Step 1.} Carry out the lightcone cuts method.}
\vskip.5\baselineskip
\noindent
Here we assume that we have obtained the future cuts, for example by using the bulk-point singularity, as 
\begin{align}
    C^+(\lambda):\quad t=T+L\cos^{-1}\qty(Y\cos\sdist{d-1}{\theta}{\Theta}),
        \label{eq: cuts in G}
\end{align}
where $\lambda = (T,Y,\bm \Theta)$ are $d+1$ parameters which we regard as the bulk coordinate.
From the general property \ref{property: injection} of cuts in \S\ref{subsubsec: review of cuts} and the above expression, $\lambda$ can move the maximum range such that the map $\lambda\mapsto C^+(\lambda)$ is injective:
\begin{align}
	T \in \mathbb R,\quad
	0\le Y\le 1,\quad
	0\le \Theta^i\le \pi ~(i=1,\cdots,d-2),\quad
	0\le \Theta^{d-1}<2\pi.
\end{align}

Because it is difficult to solve the tangential conditions \eqref{eq: tangentcuts}, we use the mathematical induction on $d$ and the symmetry of \eqref{eq: cuts in G}.
Leaving the detail of the computation of the conformal metric to appendix \ref{app: calculation detail}, here we describe the outline:
\begin{enumerate}
    \renewcommand{\labelenumi}{1-\alph{enumi})}
    \item Reconstruct the conformal metric in $d=2$.
    \item Assuming that the conformal metric in $d=D-1$ $(D\ge 3)$ has already been reconstructed, reconstruct the conformal metric in $d=D$ on the equator ($\Theta^i=\pi/2$).
    \item Rotate the coordinate in the above result to obtain the conformal metric for any $\bm \Theta$.
\end{enumerate}
Following these three steps gives us the conformal metric as
\begin{align}
    ds^2 = e^{2\omega(\lambda)} \left(-(1-Y^2)dT^2+L^2dY^2+L^2Y^2(1-Y^2)d\Omega_{d-1}^2 \right),\label{eq: conformal metric in G}
\end{align}
where $\omega(\lambda)$ is the undetermined conformal factor.
\newpage

\vskip\baselineskip
\noindent
{\it \underline{Step 2.} Build CISs from the lightcone cuts and obtain the minimal surfaces.}
\vskip.5\baselineskip
\noindent
Since we are now supposing a CFT in its ground state, the bulk must be static and rotationally symmetric.
Therefore, we can make an ansatz $\omega = \omega(Y)$ and only consider the CIS for a ball $A$ of radius $\alpha$ centered at the north pole on the time slice $T=0$.
From \eqref{eq: static CIS} with $\bm x_0$ set to zero, it is written as
\begin{align}
    \mathcal{C}_A:\quad Y=\frac{\cos(\alpha/L)}{\cos\Theta^1},\quad\mbox{or equivalently,}\quad
    \Theta^1 = \cos^{-1}\left(Y^{-1}\cos\left(\frac{\alpha}{L} \right) \right).
        \label{eq: CIS in G}
\end{align}
On the other hand, $\mathcal{C}_{\bar{A}}$ ($\bar A$ is the ball of radius $\pi-\alpha$ centered at the south pole) is given by 
\begin{align}
    \mathcal{C}_{\bar{A}}:\quad Y&=\frac{\cos((\pi L-\alpha)/L)}{\cos(\pi - \Theta^1)}
    =\frac{\cos(\alpha/L)}{\cos\Theta^1}.
\end{align}
Now, we see $\mathcal C_A = \mathcal C_{\bar A}$, thus we conclude $\mathcal E_A = \mathcal C_A$.

\vskip\baselineskip
\noindent
{\it \underline{Step 3.} Identify the conformal factor with the Ryu-Takayanagi formula.}
\vskip.5\baselineskip
\noindent
We have proven that $\mathcal E_A$ is also given as \eqref{eq: CIS in G}.
Plugging \eqref{eq: conformal metric in G} and the second expression of \eqref{eq: CIS in G} into the extremal condition of the area, with the ansatz $\omega = \omega(Y)$, we obtain
\begin{align}
    (1-Y^2)\,\omega'(Y)-2Y=0,\quad \ie, \quad  \omega(Y)=\ln\frac{C}{1-Y^2}.
        \label{eq: equation for omega in G}
\end{align}
Then the metric we now have is
\begin{align}
	ds^2 = C\left[-\frac{dT^2}{1-Y^2}+\frac{L^2 dY^2}{(1-Y^2)^2}+\frac{L^2Y^2}{1-Y^2}d\Omega_{d-1}^2 \right],
	\label{eq: rigid conformal metric}
\end{align}
which will become the global AdS$_{d+1}$ of radius $L$ after a suitable coordinate transformation, if $C=1$.

To determine $C$, we use the Ryu-Takayanagi formula \eqref{eq: RTformula}.
We introduce the cutoff as $Y = 1-\epsilon^2/L^2$ so that the metric near the boundary behaves as $ds^2\sim (L^2/\epsilon^2)ds^2_\mathrm{bdy}$.
With this cutoff, one finds that the area of $\mathcal E_A$ is reduced to
\begin{align}
	\mathrm{Area}\left(\mathcal E_A \right)=\frac{\Omega_{d-2}\,(LC)^{d-1}}{4G}\int_1^{L^2\sin(\alpha/L)/\sqrt{2L^2\epsilon^2-\epsilon^4}}\dd{\xi}(\xi^2-1)^{(d-3)/2}.
	\label{eq: Area}
\end{align}
The universal term can be extracted easily and comparing it with \eqref{eq: entanglement in G} gives $C=1$.

\subsubsection{Poincar\'e patch}\label{subsubsec: reconstruct P}
Let us suppose the ground state of a holographic CFT on $\mathbb{R}^{1,d-1}$,
\begin{align}
    ds^2_\text{bdy}=-dt^2+d\bm{x}^2,
        \label{eq: bdyg in P}
\end{align}
and that the universal term of the entanglement entropy is given as \eqref{eq: entanglement in G}.
\newpage

\vskip\baselineskip
\noindent
{\it \underline{Step 1.} Carry out the lightcone cuts method.}
\vskip.5\baselineskip
\noindent
We start from assuming that the future cuts have been obtained as 
\begin{align}
    t=T+\sqrt{(\bm{x}-\bm{X})^2+Z^2},
        \label{eq: cuts in P}
\end{align}
where we regard $\lambda = (T,Z,\bm X)$ as the coordinate of the bulk.
Then, the expression implies $T, X^i\in \mathbb R$ and $Z\ge 0$.
Solving the tangential conditions \eqref{eq: tangentcuts} gives a family of null vectors:
\begin{align}
    \delta \lambda(\lambda;\bm x) =\sqrt{(\bm{x}-\bm{X})^2+Z^2}\,\partial_T-Z\,\partial_Z+(\bm{x}-\bm{X})\vdot\partial_{\bm{X}}.
\end{align}
The condition \eqref{eq: homogeneous equation} has to hold for any $\bm x$, from which we find\footnote{
For example in determining $g_{\mu\nu}$ for $\mu,\nu = T,Y,X^1$, we can set $x^i = X^i\,(i\neq 1)$.
Then the condition, $\delta\lambda^\mu \delta\lambda^\nu g_{\mu\nu} = 0\,(\forall x^1)$, becomes that in $d=2$, which is easily solved by Tayler-expanding the l.h.s.\ for $x^1$.
}
\begin{align}
	g_{ZZ}=g_{X^iX^i}=-g_{TT},\qquad
	g_{TZ}=g_{TX^i}=g_{ZX^i}=0.
\end{align}
Therefore, we obtain the conformal metric as
\begin{align}
    ds^2= e^{2\omega(\lambda)}\left(-dT^2+dZ^2+d\bm{X}^2 \right).
\end{align}

\vskip\baselineskip
\noindent
{\it \underline{Step 2.} Build CISs from the lightcone cuts and obtain the minimal surfaces.}
\vskip.5\baselineskip
\noindent
Since we can make an ansatz $\omega = \omega(Z)$ for the same reason as before, we only consider the CIS for a ball $A$ of radius $\alpha$ centered at $(T,\bm x)= (0,\bm{0})$: 
\begin{align}
    \mathcal{C}_A:\quad |\bm X| = \sqrt{\alpha^2-Z^2},
        \label{eq: CIS in P}
\end{align}
In this case, we cannot use the argument $\mathcal{C}_A=\mathcal{C}_{\bar{A}}\Rightarrow\mathcal{E}_A=\mathcal{C}_A$, since $\mathcal{C}_A$ and $\mathcal{C}_{\bar{A}}$ never coincide.
We here assume that $\mathcal{E}_A=\mathcal{C}_A$ can be shown by using some way like we have explained in \S\ref{subsec: surfaces}.

\vskip\baselineskip
\noindent
{\it \underline{Step 3.} Identify the conformal factor with the Ryu-Takayanagi formula.}
\vskip.5\baselineskip
\noindent
Since the minimal surface has been obtained as \eqref{eq: CIS in P}, substituting it into the extremal condition of area gives
\begin{align}
    \omega'(Z)Z+1=0,\qquad \ie,\qquad
     \omega(Z)=\ln\frac{C}{Z}.
\end{align}
With cutoff $Z=\epsilon$, the area of $\mathcal E_A$ is given as \eqref{eq: Area}, but with the upper limit replaced by $\alpha/\epsilon$.
Therefore, $C$ is determined by comparing the area with \eqref{eq: entanglement in G} (the inside of the log in \eqref{eq: even S_A} is replaced with $\alpha/L$, but the coefficient is the same), which gives
\begin{align}
	ds^2 = \frac{L^2}{Z^2}\left(-dT^2 + dZ^2 +d \bm X^2 \right).
\end{align}

\subsubsection{Rindler patch}\label{subsubsec: reconstruct R}
Finally, we consider the thermal state of a holographic CFT on $\mathbb{R}\times\mathbb{H}^{d-1}$,
\begin{align}
    ds^2_\text{bdy}=-dt^2+L^2 dH_{d-1}^2.
        \label{eq: bdyg in R}
\end{align}
The universal term in the entanglement entropy is again supposed to be \eqref{eq: entanglement in G}.

\vskip\baselineskip
\noindent
{\it \underline{Step 1.} Carry out the lightcone cuts method.}
\vskip.5\baselineskip
\noindent
Let us assume that we have obtained the future cuts as
\begin{align}
    t=T+L\cosh^{-1}\qty(Y\cosh\hdist{d-1}{\chi}{X}).
        \label{eq: cuts in R}
\end{align}
From this, we can determine the range where $\lambda = (T,Y,\bm X)$ runs:
\begin{align}
	T\in \mathbb R,\quad
	Y\geq 1,\quad
	X^1\in \mathbb R,\quad
	0\le X^i\le \pi~(i=2,\cdots d-2),\quad
	0\le X^{d-1} <2\pi.
\end{align}
The computation of the conformal metric is parallel to the case of the global AdS (see appendix \ref{app: calculation detail}), so by following it, one can find
\begin{align}
    ds^2=e^{2\omega(\lambda)}\left(-(Y^2-1)dT^2+L^2dY^2+L^2Y^2(Y^2-1)dH_{d-1}^2 \right).
\end{align}

\vskip\baselineskip
\noindent
{\it \underline{Step 2.} Build CISs from the lightcone cuts and obtain the minimal surfaces.}
\vskip.5\baselineskip
\noindent
Since the state is static and $\mathrm{SO}(1,d-1)$-invariant, we can make an ansatz $\omega = \omega(Y)$.
Thus, the CIS for a ball $A$ of radius $\alpha$ centered at $(T,\bm X)=(0,\bm{0})$ is sufficient to reconstruct $\omega$:
\begin{align}
    \mathcal{C}_A:\quad Y=\frac{\cosh(\alpha/L)}{\cosh X^1}.
        \label{eq: CIS in R}
\end{align}
We require $\mathcal{E}_A=\mathcal{C}_A$ as we did above.

\vskip\baselineskip
\noindent
{\it \underline{Step 3.} Identify the conformal factor with the Ryu-Takayanagi formula.}
\vskip.5\baselineskip
\noindent
Since the minimal surface \eqref{eq: CIS in R} is obtained as $X^1=X^1(Y)$ in step 2, substituting the expression into the extremal condition provides the same differential equation for $\omega$ as in \eqref{eq: equation for omega in G}, except the range of $Y$.
 Thus the solution is
\begin{align}
    \omega(Y)=\ln\frac{C}{Y^2-1}.
        \label{eq: Omega in R}
\end{align}
Putting cutoff $Y=1+\epsilon^2/L^2$, one again finds that the area of $\mathcal E_A$ is given as \eqref{eq: Area}, with the upper limit suitably replaced.
We can fix $C$ by comparing the area with \eqref{eq: entanglement in G}, from which we conclude
\begin{align}
    ds^2=-\frac{1}{Y^2-1}dT^2+\frac{L^2}{(Y^2-1)^2}dY^2+\frac{L^2Y^2}{Y^2-1}dH_{d-1}^2.
\end{align}
It is easy to see that this is the Rindler AdS$_{d+1}$ of radius $L$.

\section{Kinematic space for higher-dimensions}\label{sec: kinematic space}
In section \ref{sec: reconstruction}, we have assumed that the set of cuts can be obtained from the boundary in some way.
In this section, we address this point by developing a candidate of higher-dimensional version of the kinematic space.
First, in \S\ref{subsec: 3D kinematic}, we review the kinematic space $K_2$ in the AdS$_3$/CFT$_2$ \cite{Czech:2019hdd}, then show how cuts are related to the geodesics on $K_2$.
After that, we generalize the framework to higher dimensions.

\subsection{Kinematic space in AdS\texorpdfstring{$_{3}$}{TEXT}/CFT\texorpdfstring{$_{2}$}{TEXT}}\label{subsec: 3D kinematic}
In AdS$_3$/CFT$_2$, the kinematic space $K_2$ is the space of all inextensible geodesics on a time slice in the bulk.
A suitable metric can be introduced on $K_2$, and thanks to that, we can reformulate the hole-ography in terms of $K_2$.
One of the benefits is that the geodesics on $K_2$ can be understood as a definition of a bulk point.
Although the original work only focused on the global patch, we will see that this interpretation is also available for the Poincar\'e patch.

\subsubsection{Definition and relation with bulk geometry}
\begin{table}[t]
    \centering
    \begin{tabular}{c|c}
        bulk AdS$_3$ & kinematic space\\ \hline\hline
        geodesic & point\\
        curve & codimension-0 region\\
        point & point-curve
    \end{tabular}
    \caption{The correspondence of the bulk of AdS$_3$ and its kinematic space $K_2$.}
    \label{tab: bulk/kinematic correspondence in AdS_3}
\end{table}

As a concrete example, let us first consider the pure global AdS$_3$:
\begin{align}
    ds^2=-\frac{r^2+L^2}{L^2}dt^2+\frac{L^2}{r^2+L^2}dr^2+r^2d\theta^2.
\end{align}
The spacelike geodesics $r=r(\theta)$ on $t=\mathrm{const.}$ are expressed as 
\begin{align}
    r(\theta)^2=\frac{L^2\cos^2(\alpha/L)}{\cos^2\qty(\theta-\phi)-\cos^2(\alpha/L)}.
        \label{eq: geodesic in global AdS_3}
\end{align}
Here $\alpha\in (0,\pi]$ and $\phi\in [0,2\pi)$ are constants characterizing the geodesic, and the map $(\alpha,\phi)\mapsto r(\theta)$ is injective.
Thus the kinematic space $K_2$, which is a space of geodesics on the time slice, is regarded as the space of $(\alpha,\phi)$.

\begin{figure}
    \centering
    \includegraphics[height=5cm]{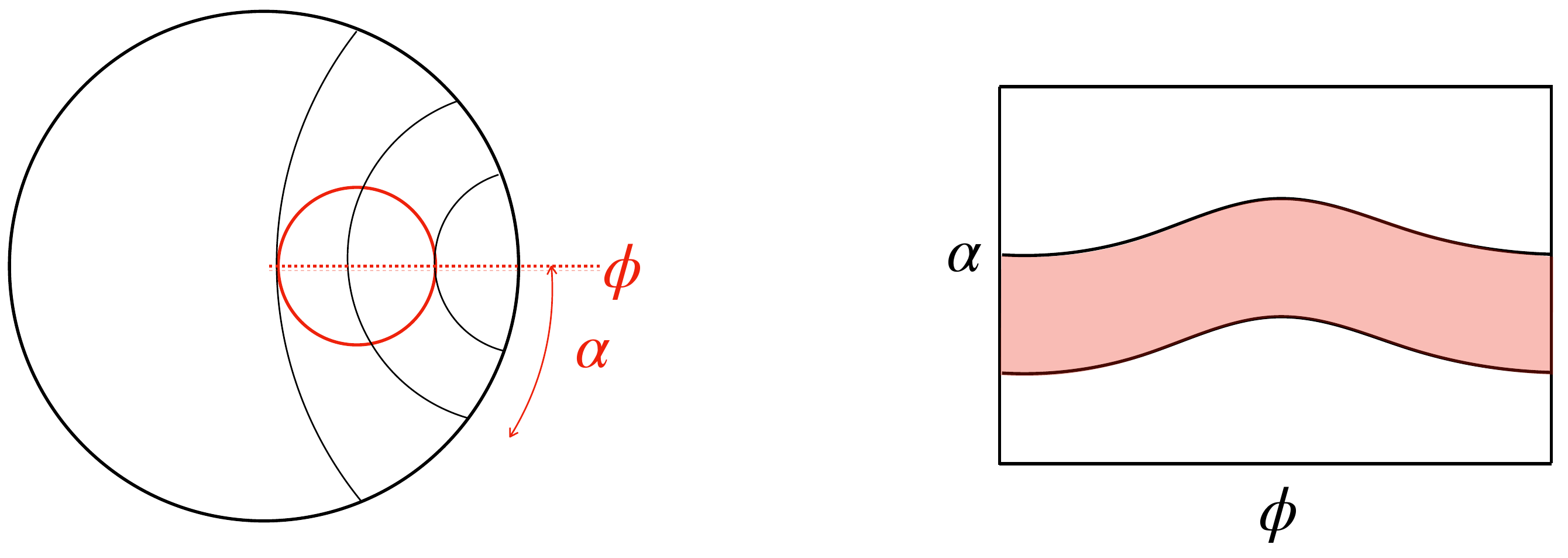}
    \caption{The region on $K_2$ that corresponds to a bulk curve.
    The left panel shows a convex curve in the bulk, and the right panel the corresponding region on $K_2$.}
    \label{fig: curve}
\end{figure}

Geometric objects on $K_2$ have bulk interpretations (Table \ref{tab: bulk/kinematic correspondence in AdS_3}).
First of all, a point on $K_2$ corresponds to the bulk geodesic by definition.
Next, let us consider a bulk curve (Fig.\,\ref{fig: curve}).
Given a closed convex curve $\gamma$ on $t=\mathrm{const.}$, we can collect all geodesics which intersect twice with $\gamma$.
The collection of such geodesics draws a codimension-0 region between two curves on $K_2$, which conversely defines a closed convex curve in the bulk.\footnote{
We can also treat open or concave curves, but it is complicated because we also have to consider how many times geodesics intersect with them.
}
Finally, collecting geodesics passing through a given bulk point draws a single curve on $K_2$; this can be regarded as the shrinking limit of a closed curve.
Putting $(r,\theta)=(R,\Theta)$ in \eqref{eq: geodesic in global AdS_3} and seeing it as a constraint on $(\alpha,\phi)$, one easily obtains such a curve on $K_2$ as
\begin{align}
    \alpha=L\cos^{-1}\qty(\frac{R}{\sqrt{R^2+L^2}}\cos(\Theta-\phi)).
        \label{eq: PC in pure global AdS}
\end{align}
We call this ``point-curve".

\subsubsection{The metric on $K_2$ and its holographic interpretation}
Let us introduce a metric to $K_2$ as 
\begin{align}
    ds_K^2=-\frac{d^2S(\alpha)}{d\alpha^2}\qty(-d\alpha^2+L^2d\phi^2),
        \label{eq: kinematic metric in d=2}
\end{align}
where $S(\alpha)$ is the area of the minimal surface for interval $[\theta-\alpha,\theta+\alpha]$ (since we are now focusing on the global AdS$_3$, it only depends on $\alpha$).
The metric was introduced so that the volume of the region on $K_2$ corresponding to a bulk curve reproduces its length --- this is the differential entropy formula.
Thanks to the Ryu-Takayanagi formula, in AdS$_3$/CFT$_2$, the $S(\alpha)$ appearing above is replaced with the entanglement entropy\footnote{
$-S''(\alpha)$ is always positive, owing to the strong subadditivity of the entanglement entropy.
In addition, it was shown that the line element is a conditional mutual information among infinitesimally different regions \cite{Czech:2015qta}.
}:
\begin{align}
    S(\alpha)=\frac{c}{3}\ln\qty(\frac{L\sin(\alpha/L)}{\varepsilon}).
\end{align}
Since we have rewritten the kinematic space in the boundary language, it now holographically provides a definition of bulk curves and their length.

Interestingly, point-curves \eqref{eq: PC in pure global AdS} can be obtained by solving the geodesic equation with respect to \eqref{eq: kinematic metric in d=2}:
\begin{align}
    \alpha'(\phi)^2-L^2+L\tan\qty(\frac{\alpha(\phi)}{L})\alpha''(\phi)=0.
\end{align}
We can holographically define the points on the time slice by using this fact, which is one of the starting points in the metric reconstruction.
The integral constants are interpreted as the coordinate of the bulk.
In the reconstruction strategy of \cite{Takeda:2021dsl}, we first obtain point-curves, which provide cuts with the help of $\mathcal C_A = \mathcal E_A$.
After reconstructing the conformal metric, the remaining conformal factor is identified by using the holographic definition of lengths introduced above.
The method can be applied to the holographic theories dual to the pure global AdS$_3$, that with conical defect, the BTZ black hole, and the AdS$_3$ soliton, as knwon examples.

All examples above correspond to the states of CFTs on $\mathbb{R}\times\mathbb{S}^1$.
Here let us check if the geodesic equation on $K_2$ to define the bulk points still works for the case of the non-compact topology.
We first consider the ground state of a holographic CFT on $\mathbb R^{1,1}$.
Since the entanglement entropy for a segment of length $\alpha$ is given by
\begin{align}
    S(\alpha)=\frac{c}{3}\ln\qty(\frac{\alpha}{\varepsilon}),
\end{align}
the metric on $K_2 $ \eqref{eq: kinematic metric in d=2} becomes
\begin{align}
    ds_K^2=\frac{c}{3\alpha^2}\qty(-d\alpha^2+dx^2).
        \label{eq: kinematic metric of pure Poincare AdS_3}
\end{align}
Then its geodesic equation is given by 
\begin{align}
    \alpha'(x)^2-1+\alpha(x)\alpha''(x)=0
\end{align}
and the general solution reads 
\begin{align}
    \alpha(x)=\sqrt{(x-\lambda^1)^2+\lambda^2}.
\end{align}
One can easily check that this describes the family of geodesics passing through $(x,z)=(\lambda^1,\lambda^2)$, where $(x,z)$ is the standard coordinate of the Poincar\'e AdS$_3$.
Since we know $\mathcal{C}_A=\mathcal{E}_A$ in pure AdS, the lightcone cuts of the bulk geometry are obtained as 
\begin{align}
    C(x;\lambda)=\lambda^0\pm\sqrt{(x-\lambda^1)^2+\lambda^2}.
\end{align}
This is consistent with \eqref{eq: cuts in P}.

We can also consider on $\mathbb R^{1,1}$ the thermal state dual to the BTZ black hole with its horizon non-compact (this includes the Rindler AdS$_3$ when the horizon radius is equal to $L$).
However, the story is the same as  the case of $\mathbb R\times \mathbb S^1$.
Note that since the boundary is non-compact, the differential entropy formula is not valid (lengths in the bulk predicted by $K_2$ will always diverge).
Nevertheless, the holographic definition of points by $K_2$ still works.

\subsection{Developing kinematic space for AdS\texorpdfstring{$_{d+1}$}{TEXT}}\label{subsec: our kinematic}
In this subsection, we extend the kinematic space to AdS$_{d+1}$ spacetimes, focusing on the question as to whether we can generalize the notion of the point-curve, in order to define bulk points in terms of the kinematic space even in higher-dimensions.

\subsubsection{Definition and relation with bulk geometry}
First, we develop a dictionary between the geometric objects similar to Table \ref{tab: bulk/kinematic correspondence in AdS_3}.
As a concrete example, let us consider the time slice $t=0$ in the global AdS$_{d+1}$,
\begin{align}
    ds^2=\frac{L^2}{r^2+L^2}dr^2+r^2d\Omega_{d-1}^2.
    \label{eq: global metric}
\end{align}
As a generalization of the geodesics in AdS$_3$ \eqref{eq: geodesic in global AdS_3}, we consider the minimal surfaces for balls.
Since we know that in the pure AdS$_{d+1}$, the minimal surface for a ball of radius $\alpha$ centered at $\bm \theta = \bm \phi$ is nothing but the corresponding CIS, it is given as
\begin{align}
    r(\bm{\theta})^2=\frac{L^2\cos^2(\alpha/L)}{\cos^2\sdist{d-1}{\theta}{\phi}-\cos^2(\alpha/L)}.
        \label{eq: minimal surfaces in global AdS_d+1}
\end{align}

If we consider the space of minimal surfaces for all boundary regions, then the dimension of the space will become uncountably infinite.
However, if we restrict to ball regions, then the dimension is $d$; $d-1$ parameters for the center and $1$ parameter for the radius.
Thus, we define a new kinematic space denoted by $K_d$ as the space of all minimal surfaces for balls, where we introduce a coordinate $(\alpha,\bm \phi)$ running over the following region:
\begin{align}
	0< \alpha\le \pi,\quad
	0\le \phi^i \le \pi ~(i=1, \cdots, d-2),\quad
	0\le \phi^{d-1} < 2\pi.
 \label{eq: K-parameter regioin in G}
\end{align}
Since now the boundary state is pure, we know $\mathcal{E}_A = \mathcal{E}_{\bar A}$, which says that \eqref{eq: K-parameter regioin in G} is redundant: both $(\alpha,\bm \phi)$ and $(\pi -\alpha,\bm \phi_a)$, where $\bm \phi_a$ is the antipodal point of $\bm \phi$, mean the identical minimal surface in the bulk.
However, the generalization to mixed states being considered, this range is suitable and we will later see that this extension is also helpful when we consider a suitable boundary condition for point-surfaces.

To define a bulk point $(R,\bm \Theta)$ in terms of $K_d$, we find the set of balls whose minimal surfaces in the bulk pass through $(R,\bm \Theta)$ as we did in $d=2$.
This constraint for balls draws codimension-1 region on $K_d$, which is obtained just by solving \eqref{eq: minimal surfaces in global AdS_d+1} as
\begin{align}
    \alpha(\bm \phi)=L\cos^{-1}\qty(\frac{R}{\sqrt{R^2+L^2}}\cos\sdist{d-1}{\Theta}{\phi}).
        \label{eq: PS in global AdS}
\end{align}
We call this ``point-surface".
The expression also appears in the cuts as we have seen in \eqref{eq: cuts in G}, which is due to $\mathcal E_A = \mathcal C_A$ for any ball $A$.

On the other hand, a convex codimension-2 surface can also be defined in terms of $K_d$, as a region of $(\alpha,\bm \phi)$'s whose minimal surfaces intersect with it.
Such a region exists between two codimension-1 surfaces on $K_d$.

Though we have only treated the global AdS$_{d+1}$, the similar things hold for the Poincar\'e and Rindler AdS$_{d+1}$.
Here we only show the result of the point-surfaces.
In the Poincar\'e patch, the point-surface corresponding to a bulk point $(Z,\bm X)$ can be extracted from the cuts as
\begin{align}
	\alpha(\bm x) = \sqrt{(\bm x - \bm X)^2+Z^2},\label{eq: PS in Poincare}
\end{align}
while in the Rindler patch, we have for a bulk point $(R,\bm X)$,
\begin{align}
	\alpha (\bm \chi) = L\cosh^{-1}\left(\frac{R}{\sqrt{R^2-L^2}}\cosh\hdist{d-1}{\chi}{X} \right).\label{eq: PS in Rindler}
\end{align}

\subsubsection{Bulk minimal surfaces from kinematic metric}\label{subsubsec: finding kinematic metric}
We have seen in \S\ref{subsec: 3D kinematic} that in AdS$_3$/CFT$_2$, the set of geodesics on $K_2$ provides a holographic definition of the bulk points.
Motivated by it, here we find a candidate of the metric on $K_d$ such that the point-surfaces extremize their area with respect to the metric.

Let us first focus on the global AdS$_{d+1}$ \eqref{eq: global metric}, where point-surfaces are given as \eqref{eq: PS in global AdS}.
For simplicity, we set the AdS radius $L=1$.
The isometry of the bulk must be reflected in $K_d$, and is actually so in the expression \eqref{eq: PS in global AdS}.
Since now we are trying find the metric that reproduces \eqref{eq: PS in global AdS}, we can make an ansatz on the form of the metric on $K_d$:
\begin{align}
	ds_K^2 = e^{2\omega_K(\alpha)}\left(-d\alpha^2 + d\Omega^2_{d-1} \right),\quad
	d\Omega_{d-2}^2 = (d\phi^1)^2 + \sin^2\phi^1 (d\phi^2)^2+\cdots.
\end{align}
Note that this defines the causal structure of $K$, which corresponds to the inclusion relations among the boundary ball regions: 1) if two boundary balls are timelike separated on $K_d$ with this metric, one contains the other, 2) if spacelike separated, one intersects the other, 3) and if null separated, one is inscribed in the other at the tangential point on the boundary.
This correspondence has already been studied in $d=2$ (see Fig.\,8 of \cite{Czech:2015qta}).

Because of the ansatz, we only need point-surfaces for $\bm \phi = \bm{0}$ in order to determine $\omega_K$.
Such point-surfaces take the form $\alpha = \alpha(\phi^1)$, whose areas are given as
\begin{align}
	\int d{\phi^1}d{\Omega_{d-2}}e^{(d-1)\omega_K(\alpha)}\sin^{d-2}\phi^1\sqrt{1-\alpha'(\phi^1)^2}.
\end{align}
The extremal condition of this functional is
\begin{align}
    \alpha''+\qty(1-\alpha'^2)\qty((d-1)\omega_K'(\alpha)+(d-2)\alpha' \cot \phi^1)=0.
        \label{eq: kinematic NG equation}
\end{align}
Substituting the point-surfaces \eqref{eq: PS in global AdS} with $\bm \Theta = \bm{0}$ into this gives 
\begin{align}
	 \omega_K'(\alpha) = -\frac{Y\cos \phi^1}{\sqrt{1-Y^2\cos^2\phi^1}}=-\dv{\alpha}\ln\sin\alpha,\quad
	 \ie,\quad
	 e^{2\omega_K(\alpha)} \propto \frac{1}{\sin^2\alpha}.
\end{align}
Some extra requirement is needed to fix the overall factor.

Following the same procedure for \eqref{eq: PS in Poincare} and \eqref{eq: PS in Rindler}, one reaches the following results:
\begin{align}
	\text{Global}&:\quad ds_K^2 = \frac{1}{\sin^2\alpha}\left(-d\alpha^2 + d\Omega^2_{d-1} \right),\label{eq: Kmetric in G} \\
	\text{Poincar\'e}&:\quad ds_K^2 = \frac{1}{\alpha^2}\left(-d\alpha^2 + d \bm x^2 \right),\\
	\text{Rindler}&:\quad ds_K^2 = \frac{1}{\sinh^2\alpha}\left(-d\alpha^2 + dH^2_{d-1} \right).\label{eq: Kmetric in R}
\end{align}
The conformal factors of these metrics are the same as those of $d=2$.
To use this in the bulk reconstruction, we have to write the metrics in terms of the boundary theory, but unfortunately, it is not so straightforward as the case of $d=2$ to replace the conformal factors with some boundary quantities.
We will discuss the matter in \S\ref{sec: discussion}.

\begin{figure}[t]
    \centering
    \begin{minipage}[t]{0.45\hsize}
        \centering
        \includegraphics[height = 4cm]{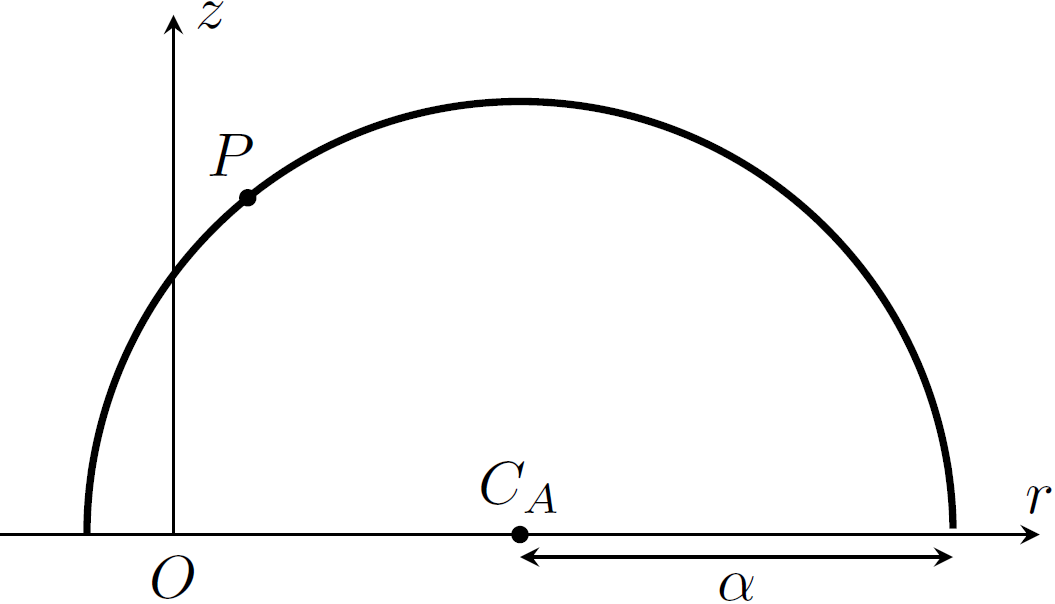}
        \subcaption{}
        \label{fig: Poincare PS}
    \end{minipage}
    \begin{minipage}[t]{0.45\hsize}
        \centering
        \includegraphics[height = 6cm]{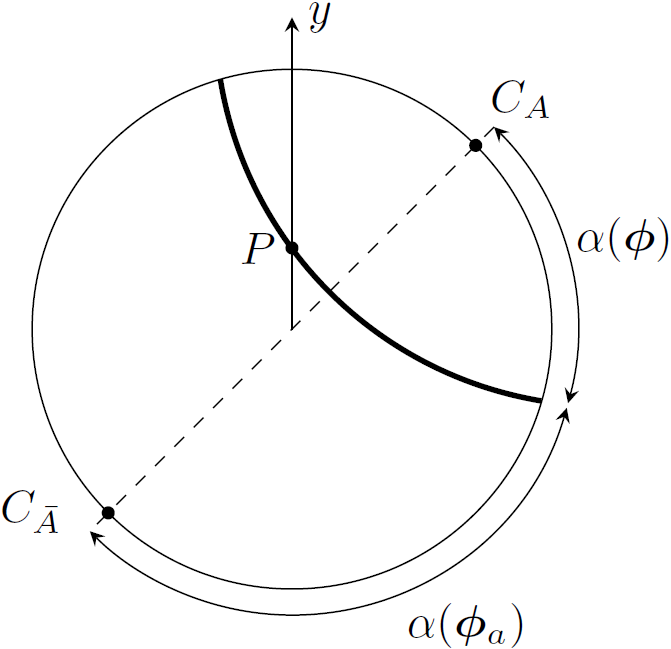}
        \subcaption{}
        \label{fig: global PS}
    \end{minipage}
    \caption{The point-surface $\alpha$ corresponding to a bulk point $P$. Each thick curve depicts a minimal surface passing through $P$, and is anchored to a boundary ball $A$ of radius $\alpha$ centered at $C_A$. (a) The Poincar\'e patch. $O$ denotes an arbitrarily chosen reference point. $\alpha$ approaches $r$ as $C_A$ goes to infinity. (b) The global patch. For boundary pure states, since $\mathcal{E}_A$ and $\mathcal{E}_{\bar{A}}$ coincide, point $(\alpha,\bm{\phi})$ in $K_d$ is identified with its antipodal point $(\pi-\alpha,\bm{\phi}_a)$ and thus \eqref{eq: bdy condition for pure G} holds.}
\end{figure}

To solve the ODE of the extremal condition and obtain point-surfaces, a certain boundary condition is needed.
In considering it, the expressions \eqref{eq: PS in global AdS}, \eqref{eq: PS in Poincare}, and \eqref{eq: PS in Rindler} are helpful.
We first consider the Poincar\'e AdS$_{d+1}$.
In \eqref{eq: PS in Poincare},  we see $\alpha\to |\bm x| + O(1)$ as $|\bm x|\to \infty$.
The physical meaning is that the bulk point approaches to the boundary when it is seen from $|\bm x|\to \infty$: $(R,\bm X)\to (\infty, \bm X)$.
This is also valid for the Rindler AdS$_{d+1}$, whose boundary is non-compact, and actually we see $\alpha\to \chi^1+O(1)$ from \eqref{eq: PS in Rindler}.
Though we have not yet surveyed the kinematic space for other geometries, the physical picture is so robust that it will be applied to other geometries with non-compact boundaries.
Thus, in a coordinate-independent way, our proposal of the boundary condition for solving the extremal condition on $K_d$ is summarized as
\begin{align}
    \alpha \to  r+O(1)\quad \text{as}\quad r\rightarrow\infty, \label{eq: bdy condition for P,R}
\end{align}
where $r$ is the geodesic distance from an arbitrarily chosen reference point (see Fig.\,\ref{fig: Poincare PS}).\footnote{
One can confirm that $\alpha(\bm x) = \sqrt{(d-2)/(d-1)}|\bm x-\bm X|$ is a solution of the extremal condition on $K_d$ of the Poincar\'{e} AdS$_{d+1}$ with $d>2$ and our proposal actually removes this kind of solutions.
}

In the global patch, where the boundary is compact, the story is different from the above.
In this case, we should rather pay attention to the periodicity of $\mathbb S^{d-1}$.
For pure states, the periodicity condition for point-surfaces must be
\begin{align}
    \alpha(\bm \phi)+\alpha(\bm \phi_a)=\pi,\qquad
    \mbox{$\bm \phi_a$ : the antipodal point of $\bm \phi$},
        \label{eq: bdy condition for pure G}
\end{align}
which is independent of \eqref{eq: kinematic NG equation}.
As depicted in Fig.\,\ref{fig: global PS}, this reflects the redundancy of \eqref{eq: K-parameter regioin in G}, and is actually satisfied by \eqref{eq: PS in global AdS}.
For mixed states in the global patch, however, $\mathcal E_A = \mathcal E_{\bar A}$ does not hold and $\mathcal E_A$ jumps at a critical size of $A$.
This jump is annoying, because the point-surfaces become discontinuous and also we cannot probe the entanglement shadows.
Instead, we adopt bulk surfaces that are non-minimum but extremal when we introduce $K_d$, just as we do in the case of the BTZ and conical AdS$_3$.
Such surfaces can wrap a black hole.
In this case, point-surfaces become multi-valued and not periodic, so we take the covering of \eqref{eq: K-parameter regioin in G} to make them single-valued.
Therefore, the boundary condition \eqref{eq: bdy condition for P,R} is also available for mixed states on $\mathbb{S}^{d-1}$.

It should be emphasized that the arguments here are based only on the general properties of point-surfaces in each patch.
Therefore, the boundary conditions we have proposed will be valid not only for the pure AdS$_{d+1}$ but also for other spacetimes.

\section{Summary and discussions}\label{sec: discussion}
We have reconstructed the global, Poincar\'e, and Rindler AdS$_{d+1}$ with given cuts, recovering the missing conformal factor in the lightcone cuts method.
In this procedure, it is found from some proposed dictionaries that the causal information surface (CIS) for any ball-shaped region is minimal.
From the set of CISs, it is able to determine the conformal factor with the Ryu-Takayanagi formula.
Our strategy also reconstructs other geometries with given cuts, whenever we can holographically detect the minimal surface for any ball from the corresponding CIS.

On the other hand, we have also proposed a new kinematic space for AdS$_{d+1}$, $K_d$, as the space of minimal surfaces for boundary balls.
The collection of boundary balls whose minimal surfaces pass a common bulk point draws a codimension-1 surface on $K_d$, which we have named point-surface.
In addition, we have introduced the metric to $K_d$, which makes point-surfaces extremize their areas.

In the following, let us discuss unsolved subjects and future directions.

\disitem{The relation between cuts and point-surfaces}\\
The set of the future/past cuts is isomorphic to the bulk region $J^\mp(\partial M)$, while the set of point-surfaces to each time slice.
Thus, for the region covered by both of them, there is a one-to-one map from the set of cuts to that of point-surfaces.
If this map is found in the boundary language, our method will give the missing conformal factor in any holographic theory, after the conformal metric is reconstructed from the cuts.
The map will also provide another possible way to obtain the cuts, if a holographic description of the kinematic space is finally found.
This has already been done for holographic theories whose bulk is locally AdS$_3$ \cite{Czech:2014ppa,Takeda:2021dsl}.

In \S\ref{subsec: surfaces}, a way to obtain CISs from cuts has been explained.
From given $\{\mathcal{C}_A|A:\mathrm{ball}\}$, inversely, each cut $C_p^\pm$ is identified as \begin{align}
    C_p^\pm = \{v_A^\pm\,|\, A\mbox{ s.t. }\mathcal{C}_A\mbox{ passes through }p\}.
\end{align}
A similar relation holds also between minimal surfaces and point-surfaces; a minimal surface is identified as the set of point-surfaces that contain a certain point $k \in K_d$.
Thus, the unknown relation between cuts and point-surfaces may be understood as that between CISs and minimal surfaces.

The relation between CISs and minimal surfaces has been studied in the viewpoints of both the bulk and boundary.
In the bulk viewpoint, the position relation between $\mathcal{C}_A$ and $\mathcal{E}_A$ has been surveyed.
Especially, it was shown in \cite{Hubeny:2012wa, Wall:2012uf, Headrick:2014cta} that, under some conditions including the null curvature condition, $\mathcal{E}_A$ must be causally disconnected from $\Diamond_A$, \ie, it must lie outside $\blacklozenge_A$.
As explained in \S\ref{sec: preparation}, the fact seems to relate to the causality of the boundary theory.
But, to accomplish a universal reconstruction through our strategy, we need to answer a stronger question: given a holographic theory, how is the position of $\mathcal{E}_A$ related to $\mathcal{C}_A$ in terms of the boundary language?
Since the dual wedge observables of $\mathcal{E}_A$ and $\mathcal{C}_A$ are considered to be the entanglement entropy \cite{Ryu:2006bv} and the one-point entropy \cite{Kelly:2013aja} respectively, studying their relation seems valuable in addressing this question.

\disitem{Speciality of pure Einstein gravity and extension}\\
Among the things related to our work, there are mainly two facts which seem peculiar to geometries solving the Einstein equation without matter: the coincidence $\mathcal{C}_A = \mathcal{E}_A$ for any ball $A$, and holographic derivation of point-surfaces (or curves).

The coincidence $\mathcal{C}_A = \mathcal{E}_A$ for ball $A$ has been confirmed for AdS$_d$, AdS$_3$ with conical singularity, AdS$_3$ soliton, and BTZ black hole both with and without the angular momentum \cite{Hubeny:2012wa,Takeda:2021dsl}.
Since all of them are locally AdS, the cause of the coincidence seemed to exist in the symmetry.
However, the evidence in appendix \ref{app: Sch-AdS} implies that it holds for less symmetric and higher-dimensional cases.
Thus, their common feature rather seems to be the fact that they solve the Einstein equation
\begin{align}
    R_{\mu\nu} - \frac{1}{2}R g_{\mu\nu} + \Lambda g_{\mu\nu}=0\qquad
    (\Lambda<0).\label{eq: pure Einstein}
\end{align}
Given that imposing some EOM can determine the conformal factor after a conformal metric is reconstructed, it may be possible that \eqref{eq: pure Einstein} picks up the conformal factor that makes CISs for balls minimal.
Studying this direction may also help us understand the question as to how $\mathcal{C}_A$ and $\mathcal{E}_A$ are related in holographic theories.

The second fact is about the kinematic space $K_d$.
A point-curve is equivalent to the collection of geodesics passing through a bulk point, and at the same time, it is itself a geodesic on $K_2$.
This fact was discovered in \cite{Czech:2014ppa} in a bottom-up way and a counterexample was found in \cite{Burda:2018rpb}.
The way we have found the suitable metrics on $K_d$ was also bottom-up, and the extension to other geometries has not been done.
Thus, the simpleness of the metrics on $K_d$ established so far could also be due to \eqref{eq: pure Einstein}, because \eqref{eq: pure Einstein} is solved by any geometry for which \eqref{eq: kinematic metric in d=2} or one of \eqref{eq: Kmetric in G} -- \eqref{eq: Kmetric in R} has been verified so far.
Therefore, in order to extend the formalism of the kinematic space to generic theories, it seems important to unravel what makes the established metrics quite simple, in particular, whether or not the fact is related to \eqref{eq: pure Einstein}.

\disitem{On the holographic redefinition of $K_d$}\\
We have not yet found a holographic description of $K_d$.
Considering the further development of the bulk reconstruction strategy with the help of the kinematic space, we first of all have to rewrite it from the boundary viewpoint.
In holographic theories whose bulk is locally AdS$_3$, the metric on $K_2$ is known to be written by the entanglement entropy as \eqref{eq: kinematic metric in d=2}, which is also understood as a conditional mutual information (CMI) among infinitesimally different regions.
Even for the higher-dimensional AdS, we guess that the origin of the distance in $K_d$ comes from the CMI.
We should confirm this expectation, though the CMI among different balls cannot so simply be computed.

Other candidate quantities related to the entanglement may also be possible in rewriting the kinematic metric.
Since the metric on $K_d$ should be independent of the cutoff appearing in the entanglement entropy, we can narrow down the list of candidates.
For example, the entanglement contour \cite{Chen_2014,Wen:2018whg,Wen:2019iyq,Rolph:2021nan}, which is a local quantity induced from the set of entanglement entropy is known to be cutoff-independent.

\disitem{The hole-ography in higher-dimensions}\\
The application of $K_d$ to the measurement of codimension-2 areas in the bulk, which is what we call hole-ography in AdS$_3$/CFT$_2$, is also interesting.
However, we can soon find that a simple analogy of the 3-dimensional case fails.
Let us consider a sphere $S$ described as $r=R$ in the global AdS$_{d+1}$, where $r$ is the radial coordinate used in \eqref{eq: global metric}.
The area of the sphere is given as $\mathrm{Area}(S) = R^{d-1}\Omega_{d-1}$ with $\Omega_{d-1}$ being the volume of the unit $(d-1)$-dimensional sphere.
In $d=2$, the area becomes the circumference, and the Crofton formula is known to work: $\mathrm{Area}(S)$ is equal to the volume of the corresponding region in $K_2$.\footnote{The formula is equivalent to the differential entropy formula, and is valid for any curves, not limited to circles \cite{Balasubramanian:2013lsa,Headrick:2014eia,Czech:2015qta}.}
A natural analogy one may expect is that the $(d-1)$-dimensional sphere in the bulk will be equal to the volume of the corresponding region in $K_d$.
From \eqref{eq: Kmetric in G}, the volume is
\begin{align}
    V(R)=\Omega_{d-1}\int_{\alpha_0}^{\pi/2}\frac{d\alpha}{\sin^{d}\alpha}~,
\end{align}
where the lower limit $\alpha_0 = \cot^{-1} R$ is determined so that the set of points on $\alpha=\alpha_0$ in $K_d$ equals the collection of minimal surfaces tangent to $S$.
Unfortunately, the $R$-dependence of $V(R)$ is different from that of $\mathrm{Area}(S)$, unless $d=2$.
Therefore, some modification or a new tool is needed for $K_d$ to be compatible with the higher-dimensional hole-ography.

\subsection*{Acknowledgement}
We thank Takuya Yoda for discussions.
The work of D.T.\ is supported by Grant-in-Aid for JSPS Fellows No.\ 22J20722.

\appendix

\section{Calculation details of reconstruction in \S\ref{subsec: explicit reconstruction}}\label{app: calculation detail}

In this appendix, we provide the detail of the calculation in step 1 of \S\ref{subsec: explicit reconstruction} to reconstruct the metric of pure AdS$_{d+1}$ spacetimes up to a conformal factor.
The calculation below is slightly technical, however, what we do is just a mathematical induction on the spacetime dimension $d$:
\begin{enumerate}
    \renewcommand{\labelenumi}{1-\alph{enumi})}
    \item Reconstruct the conformal metric in $d=2$.
    \item Assuming that the conformal metric in $d=D-1$ $(D\ge 3)$ has already been reconstructed, reconstruct the conformal metric in $d=D$ at a certain point.
    \item Obtain the conformal metric at any point by rotating the coordinate in the above result with the symmetry transformation of the lightcone cuts.
\end{enumerate}

From now on, we show the calculation in the global patch, and the application to the Rindler patch is straightforward.
\vskip\baselineskip
\noindent
{\it \underline{1-a)} Reconstruct the $d=2$ metric.}
\vskip.5\baselineskip
Although the metric of global AdS$_3$ geometry has already been reconstructed in \cite{Takeda:2021dsl}, here we dare to provide the calculation again to be self-contained.
In the $d=2$ case, from the lightcone cuts \eqref{eq: cuts in G}, the tangential conditions \eqref{eq: tangentcuts} are given as 
 \begin{align}
     0&=\delta T-\frac{L\qty(\delta Y\cos(\theta-\Theta)+Y\delta\Theta\sin(\theta-\Theta))}{\sqrt{1-Y^2\cos^2(\theta-\Theta)}},
         \label{eq: 1 in G d=2}\\
     0&=\delta Y\sin(\theta-\Theta)-Y(1-Y^2)\delta\Theta\cos(\theta-\Theta),
        \label{eq: 2 in G d=2}    
\end{align}
up to $O(\epsilon)$.
Solving these equations gives a bulk null vector 
\begin{align}
    \delta X=-LY\sqrt{1-Y^2\cos^2(\theta-\Theta)}\partial_T-Y(1-Y^2)\cos(\theta-\Theta)\partial_Y-\sin(\theta-\Theta)\partial_\Theta.
\end{align}
Then the condition \eqref{eq: homogeneous equation} becomes 
\begin{align}
    0=&g_{TT}L^2Y^2\qty(1-Y^2\cos^2(\theta-\Theta))+2g_{TY}LY^2(1-Y^2)\sqrt{1-Y^2\cos^2(\theta-\Theta)}\nonumber\\
    &+2g_{T\Theta}LY\sin(\theta-\Theta)\sqrt{1-Y^2\cos^2(\theta-\Theta)}+g_{YY}Y^2(1-Y^2)\cos^2(\theta-\Theta)\nonumber\\
    &+2g_{Y\Theta}Y(1-Y^2)\sin(\theta-\Theta)\cos(\theta-\Theta)+g_{\Theta\Theta}\sin^2(\theta-\Theta),
    \label{eq: full order in 3D}
\end{align}
and expanding it around $\theta=\Theta$ gives
\begin{align}
    0=&g_{TT}L^2Y^2(1-Y^2)+2g_{TY}LY^2(1-Y^2)^{\frac{3}{2}}-g_{YY}Y^2(1-Y^2)^2\nonumber\\
    &+\qty(2g_{T\Theta}LY\sqrt{1-Y^2}+2g_{Y\Theta}Y(1-Y^2))(\theta-\Theta)\nonumber\\
    &+\qty(g_{\Theta\Theta}+g_{TT}L^2Y^4+g_{YY}Y^2(1-Y^2)^2-g_{TY}LY^2(1-Y^2)(1-2Y^2)/\sqrt{1-Y^2})(\theta-\Theta)^2\nonumber\\
    &+\qty(-\frac{4}{3}g_{Y\Theta}Y(1-Y^2)-g_{T\Theta}\frac{LY(1-4Y^2)}{3\sqrt{1-Y^2}})(\theta-\Theta)^3\nonumber\\
    &+\frac{1}{3}\qty(-g_{\Theta\Theta}-g_{TT}L^2Y^4+g_{YY}Y^2(1-Y^2)^2+g_{TY}\frac{LY^2(1-12Y^2+8Y^4)}{4\sqrt{1-Y^2}})(\theta-\Theta)^4\nonumber\\
    &+O((\theta-\Theta)^5).
\end{align}
From the condition that the above equation has to hold for any $\theta$, we have
\begin{align}
     g_{TT}=-\frac{1-Y^2}{L^2}g_{YY},\quad
     g_{\Theta\Theta}=Y^2(1-Y^2)g_{YY},\quad
     g_{TY}=g_{T\Theta}=g_{Y\Theta}=0.
\end{align}
This is also sufficient for \eqref{eq: full order in 3D}.
Therefore, the conformal metric of the bulk geometry in the $d=2$ case is determined as 
\begin{align}
    ds^2\propto-(1-Y^2)dT^2+L^2dY^2+L^2Y^2(1-Y^2)d\Theta^2\qquad(d=2).
\end{align}
\vskip\baselineskip
\noindent
{\it \underline{1-b)} Reconstruct the $d=D$ metric in the equatorial plane.}
\vskip.5\baselineskip
In the equatorial plane ($\Theta^i=\pi/2\ (i=1,\dots,D-1)$) the tangential conditions \eqref{eq: tangentcuts} become 
\begin{align}
     0=&\delta T-\frac{\mathcal{S}_{D-1}\delta Y}{\sqrt{1-Y^2\mathcal{S}_{D-1}^2}}+\sum_{k=1}^{D-1}\frac{Y\partial_k\mathcal{S}_k}{\sqrt{1-Y^2\mathcal{S}_{D-1}^2}}
         \label{eq: 1 in G}\\
     0=&-\delta Y\cot\theta^k\mathcal{S}_{D-1}+\sum_{i=1}^{k-1}Y^3\cot\theta^k\,\partial_i\mathcal{S}_i\,\mathcal{S}_{D-1}^2\delta\Theta^i-Y\mathcal{S}_k\qty(1-\frac{Y^2}{\sin^2\theta^k}\mathcal{S}_{D-1}^2)\delta\Theta^k\nonumber\\
     &+\sum_{i=k+1}^{D-1}Y\cot\theta^k\,\partial_i\mathcal{S}_{i}\,\delta\Theta^i\qquad(k=1,\dots,D-1),
         \label{eq: 2_k in G}
\end{align}
where $\mathcal{S}_k:=\sin\theta^1\cdots\sin\theta^k,\ \mathcal{S}_0:=1$ and $\partial_k:=\partial/\partial\theta^k$.

First, when we set $\delta\Theta^{D-1}=0$ and $\theta^{D-1}=\pi/2$, the condition \eqref{eq: 2_k in G} for $k=D-1$ is trivially satisfied and the others are the same as in the $d=D-1$ case. According to the assumption of the mathematical induction that the metric of the pure global AdS$_{D-1}$ be reconstructed from the \eqref{eq: 1 in G}, \eqref{eq: 2_k in G}, the $D\times D$ components of the metric are determined as 
\begin{align}
     &g_{TT}=-\frac{1-Y^2}{L^2}g_{YY},\\
     &g_{\Theta^i\Theta^i}=Y^2(1-Y^2)g_{YY}\qquad(i=1,\dots,D-2),\\
     &g_{TY}=g_{T\Theta^i}=g_{Y\Theta^i}=g_{\Theta^i\Theta^j}=0\qquad(i,j=1,\dots,D-2,\ i\neq j).    
\end{align}

Second, when we set $\delta\Theta^i=0$ and $\theta^i=\pi/2\ (i=1,\dots,D-2)$, the nontrivial ones among the conditions \eqref{eq: 1 in G}, \eqref{eq: 2_k in G} become 
\begin{align}
     0&=\delta T-\frac{L\qty(\delta Y\sin\theta^{D-1}-Y\delta\Theta^{D-1}\cos\theta^{D-1})}{\sqrt{1-Y^2\sin^2\theta^{D-1}}},\\
     0&=\delta Y\cos\theta^{D-1}+Y(1-Y^2)\delta\Theta^{D-1}\sin\theta^{D-1}.    
\end{align}
Since this is the same as the $d=2$ case \eqref{eq: 1 in G d=2}, \eqref{eq: 2 in G d=2} with $\Theta=\pi/2$, some components of the metric are determined as 
\begin{align}
     g_{\Theta^{D-1}\Theta^{D-1}}=Y^2(1-Y^2)g_{YY},\qquad
     g_{T\Theta^{D-1}}=g_{Y\Theta^{D-1}}=0.
\end{align}

Finally, when we set $\delta\Theta^k=0,\ \theta^k=\pi/2\ (k\neq i,D-1)$ and $\theta^i=\theta^{D-1}=\pi/4$, the nontrivial ones among the conditions \eqref{eq: 1 in G}, \eqref{eq: 2_k in G} become 
\begin{align}
     &\delta T-\frac{L\qty(\delta Y-Y\qty(\sqrt{2}\delta\Theta^i+\delta\Theta^{D-1}))}{2\sqrt{1-Y^2/4}}=0,\\
     &\delta Y+\frac{Y}{\sqrt{2}}(2-Y^2)\delta\Theta^i-Y\delta\Theta^{D-1}=0,\\
     &2\delta Y-\frac{Y^3}{\sqrt{2}}\delta\Theta^i+Y(2-Y^2)\delta\Theta^{D-1}=0.    
\end{align}
Its solution reads
\begin{align}
    \delta X=-L\sqrt{4-Y^2}\partial_T-(1-Y^2)\partial_Y+\frac{\sqrt{2}}{Y}\partial_{\Theta^i}+\frac{1}{Y}\partial_{\Theta^{D-1}},
\end{align}
and substituting this into the null condition \eqref{eq: homogeneous equation} gives $g_{\Theta^i\Theta^{D-1}}=0\ (i=1,\dots,D-2)$.
Therefore, the conformal metric in the equatorial plane is determined as 
\begin{align}
    ds^2\propto-(1-Y^2)dT^2+L^2dY^2+L^2Y^2(1-Y^2)\sum_{i=1}^{D-1}\qty(d\Theta^i)^2\qquad\qty(\Theta^i=\frac{\pi}{2}).
        \label{eq: conf. metric in Theta=pi/2, G}
\end{align}
\vskip\baselineskip
\noindent
{\it \underline{1-c)} Reconstruct the metric at any $\Theta$.}
\vskip.5\baselineskip
The lightcone cuts \eqref{eq: cuts in G} have the $SO(d)$ symmetry as
\begin{align}
    C(\bm{\theta}_O;T,Y,\bm{\Theta}_O)=C(\bm{\theta};T,Y,\bm{\Theta}),
        \label{eq: cutsSO(d)}
\end{align}
where the subscript $O$ is used to express the coordinates of the points rotated by $O\in SO(d)$.
Now we use this symmetry to rotate an arbitrary point $\bm{\Theta}$ to the point $\Theta_O^i=\pi/2\ (i=1,\dots,d-1)$, which satisfies 
\begin{align}
    \sum_{i=1}^{d-1}\qty(d\Theta^i_O)^2=\sum_{i=1}^{d-1}\sin^2\Theta^1\cdots\sin^2\Theta^{i-1}\qty(d\Theta^i)^2\qty(=d\Omega_{d-1}^2).
        \label{eq: trf. of dOmega}
\end{align}
Since the cuts are unchanged under this rotation, the metric at $(T,Y,\bm{\Theta})$ is obtained just by substituting \eqref{eq: trf. of dOmega} into the metric at $(T,Y,\bm{\Theta}_O)$ of the form \eqref{eq: conf. metric in Theta=pi/2, G}: 
\begin{align}
    ds^2\propto-(1-Y^2)dT^2+L^2dY^2+L^2Y^2(1-Y^2)d\Omega_{d-1}^2.
\end{align}

\section{Confirmation of \texorpdfstring{$\mathcal{C}_A = \mathcal{E}_A$}{TEXT} for AdS black holes}\label{app: Sch-AdS}

\begin{figure}[p]
    \centering
    \begin{tabular}{ccc}
         $d=3$ & $d=4$ & $d=5$\\[24pt]
         \coinFig{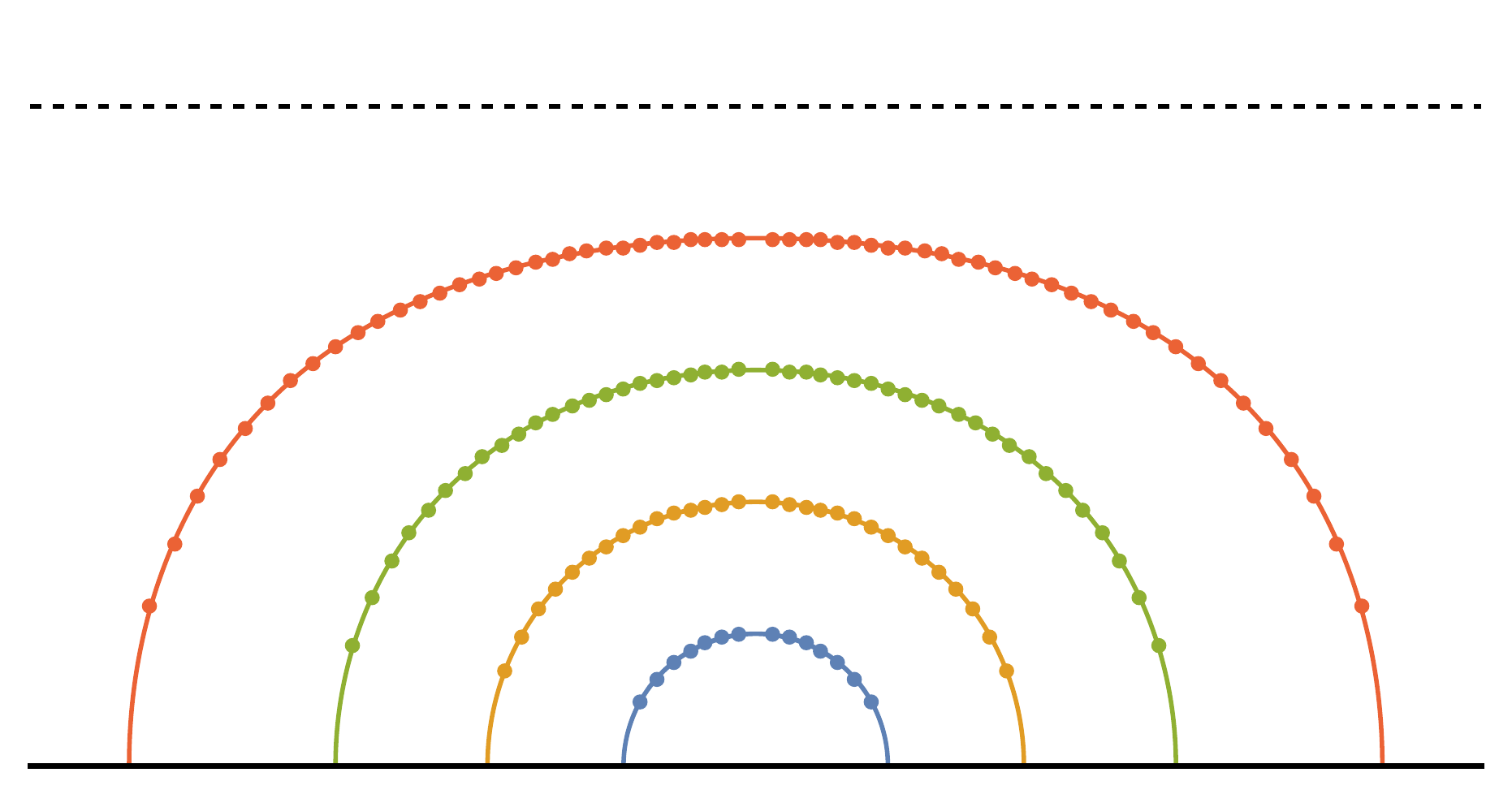} & \coinFig{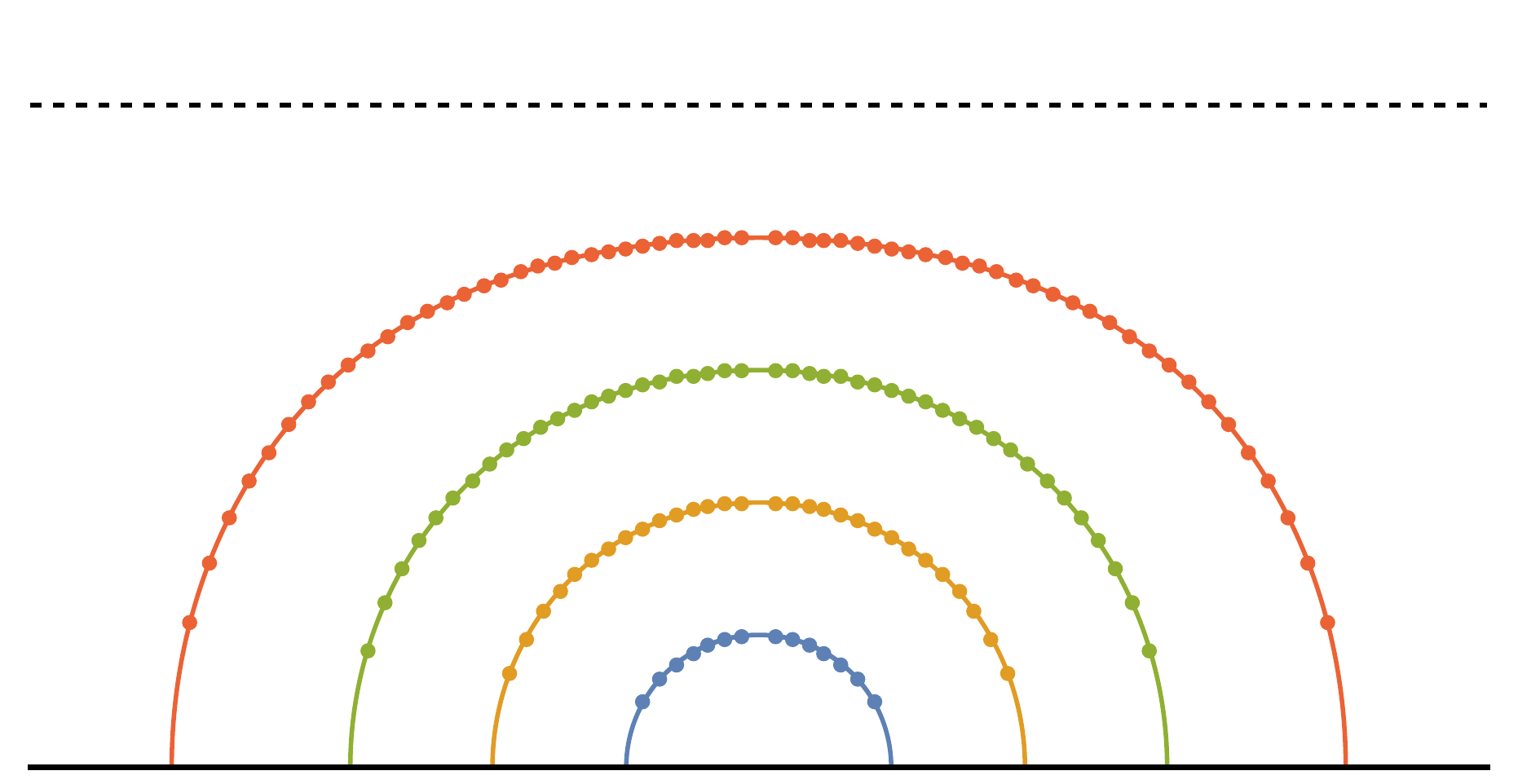} & \coinFig{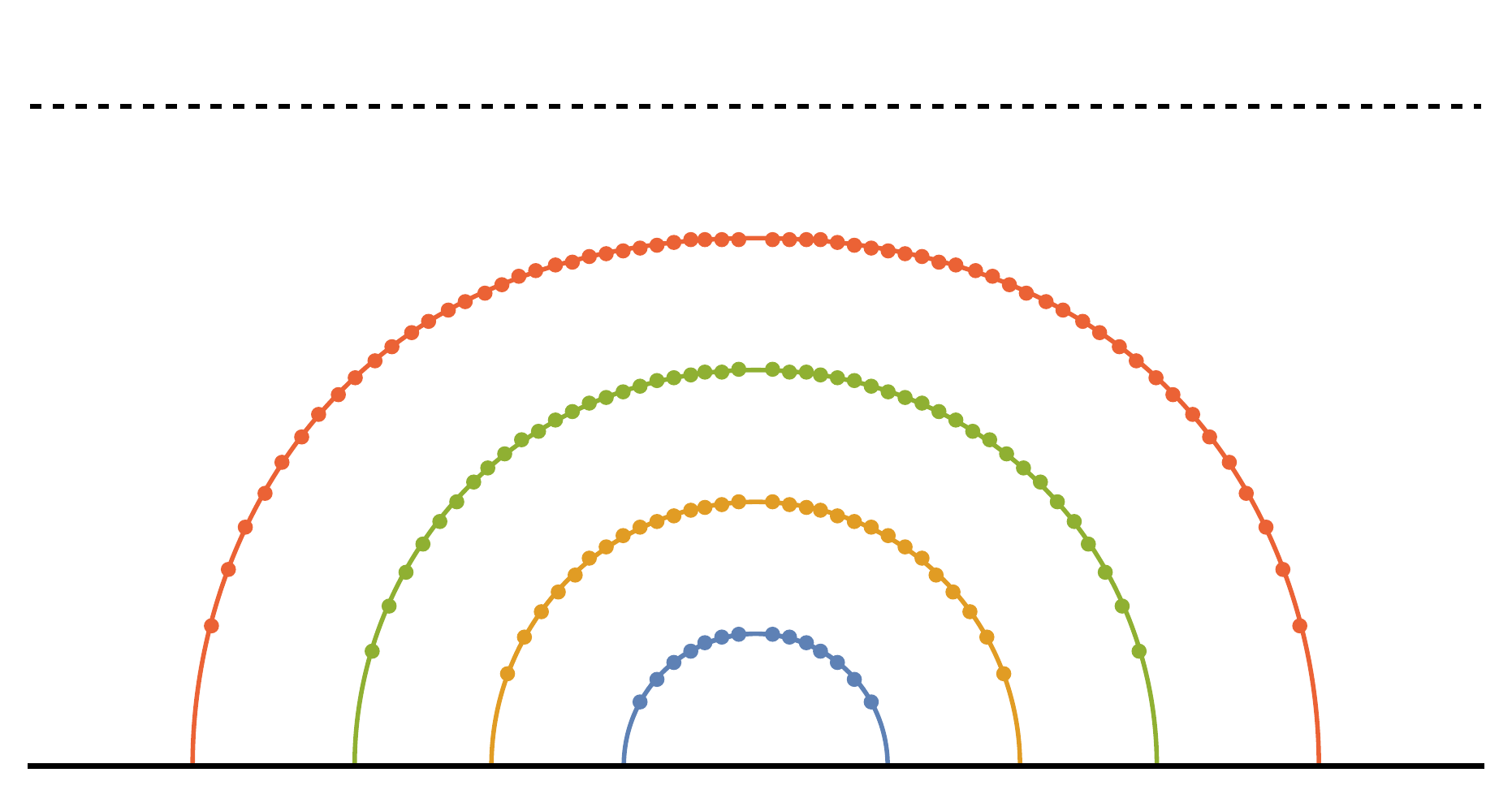}\\[24pt]
         \coinFig{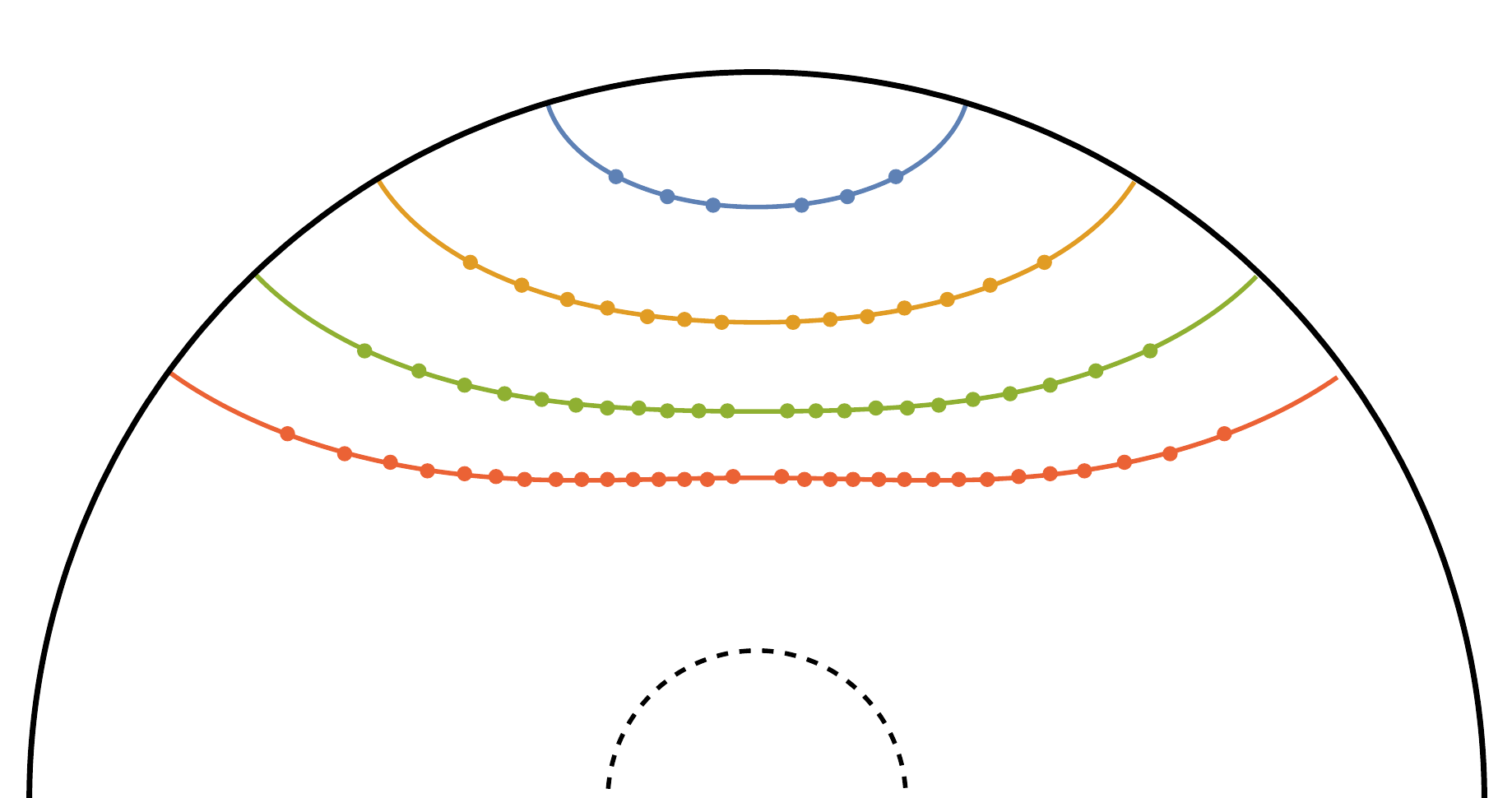} & \coinFig{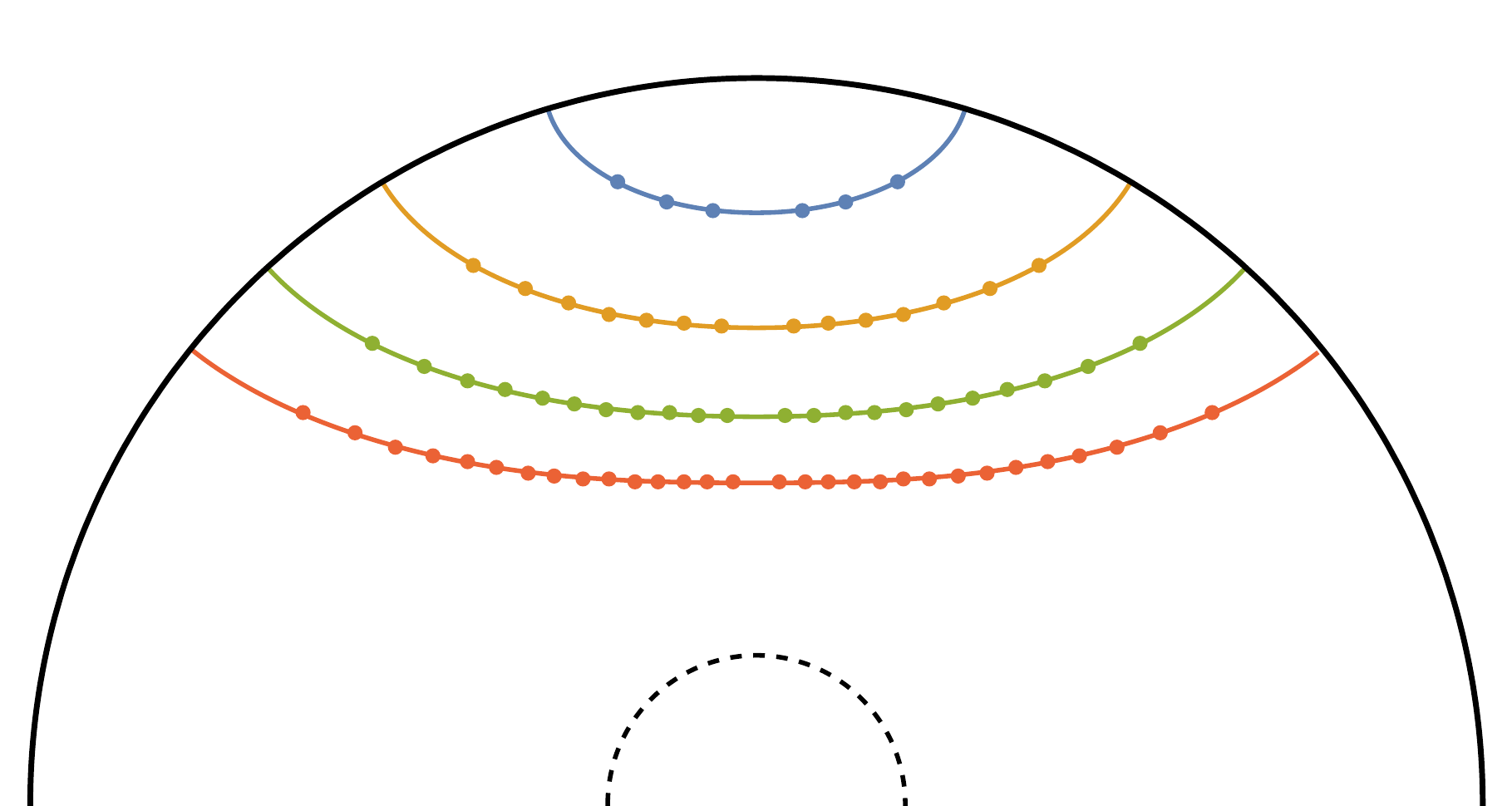} & \coinFig{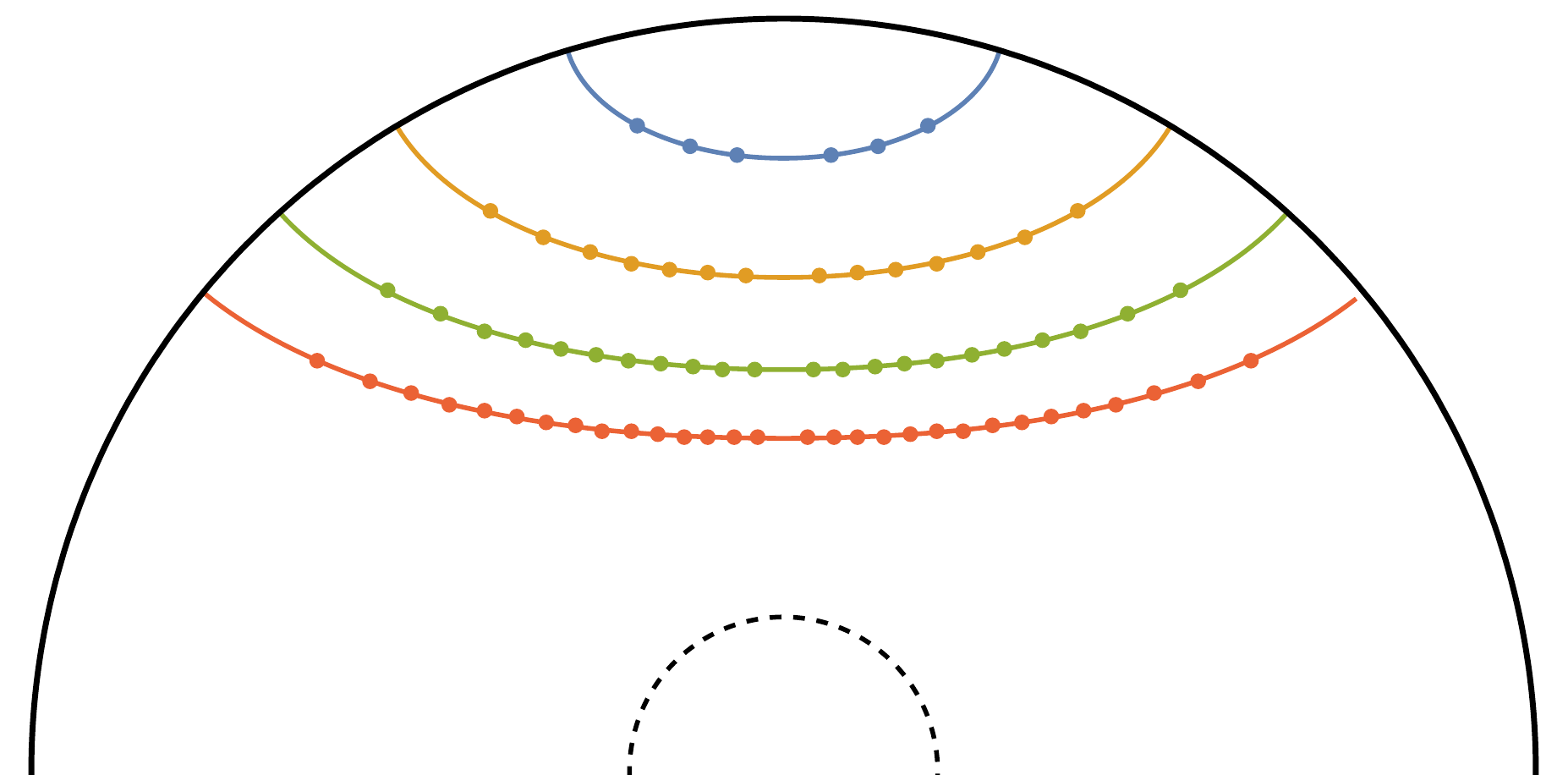}\\[24pt]
         \coinFig{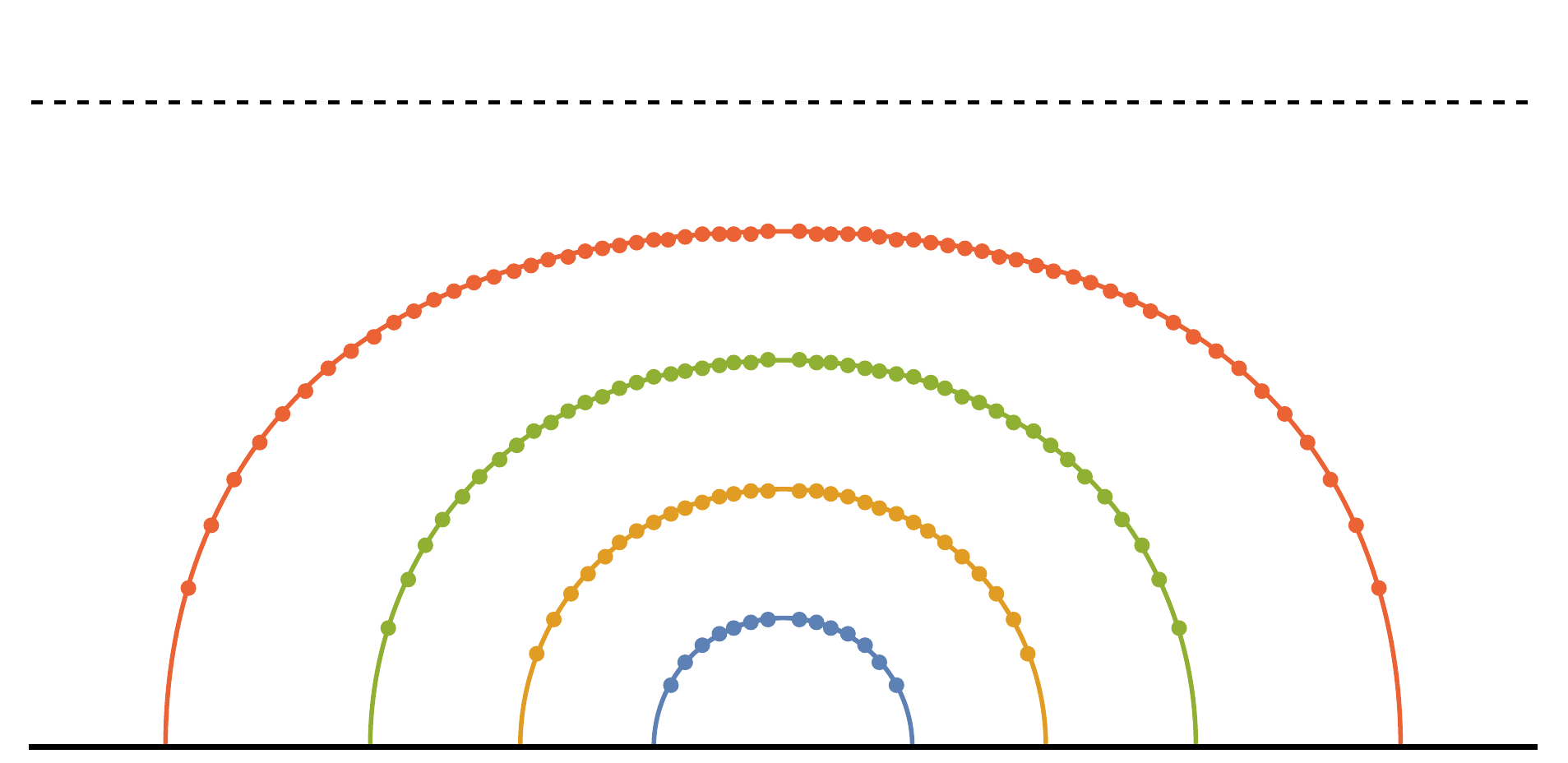} & \coinFig{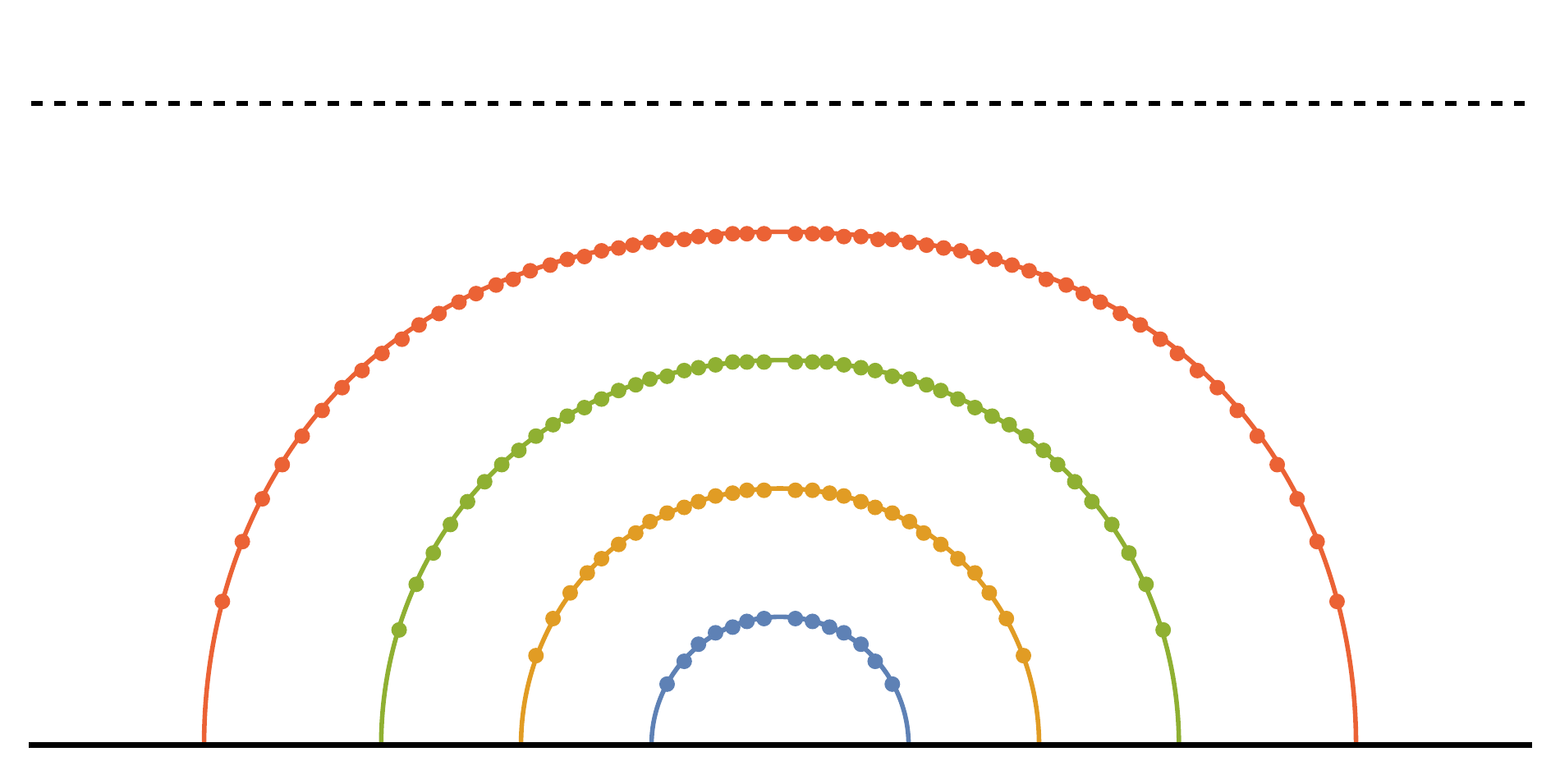} & \coinFig{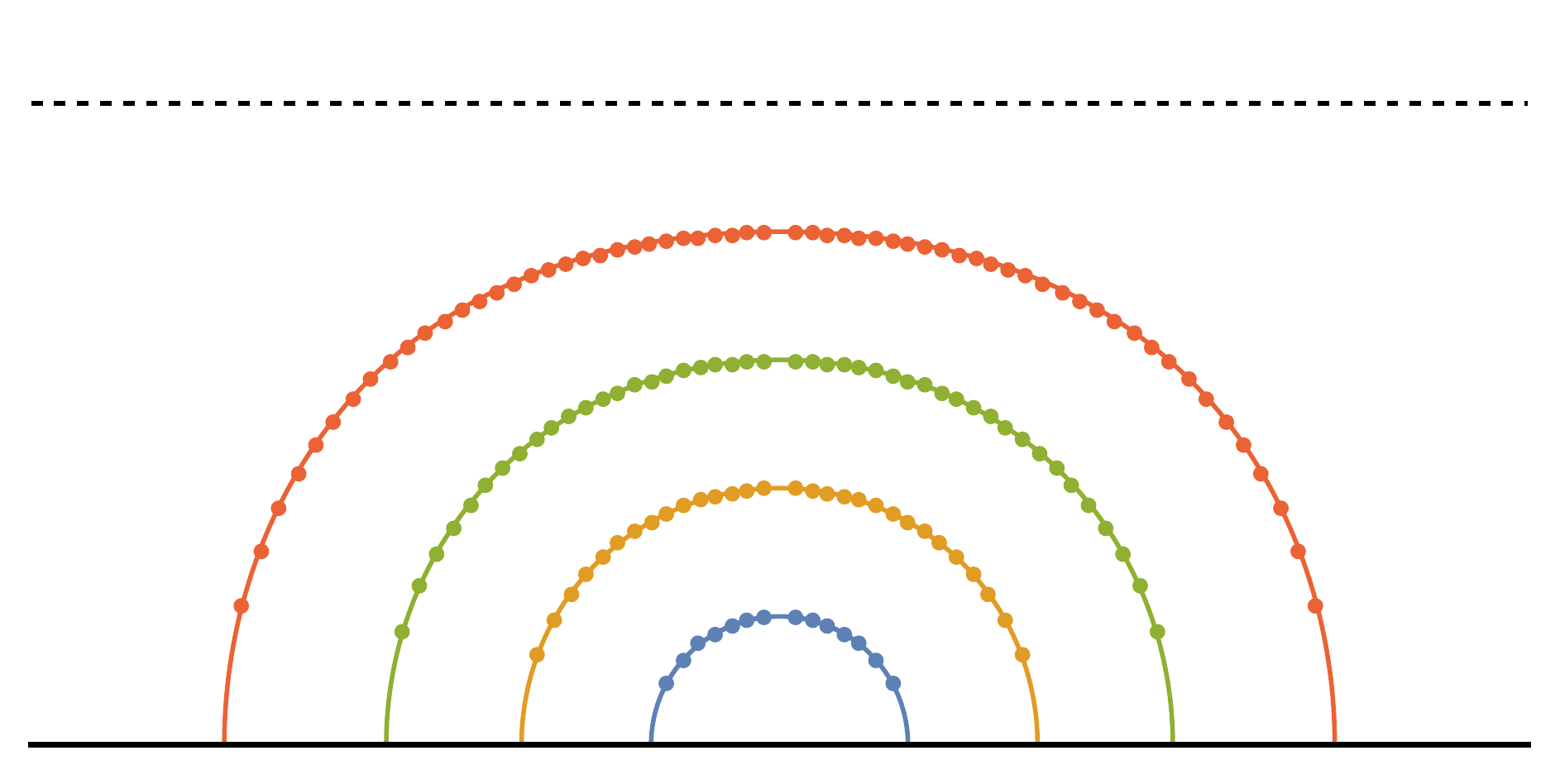}
    \end{tabular}
    \caption{Numerical check of $\mathcal{C}_A = \mathcal{E_A}$ for planar, spherical, and hyperbolic black hole geometries.
    They are aligned in this order from the top to the bottom.
    Each black line is the boundary and each dashed black line is the black hole horizons.
    The solid colored lines are minimal surfaces for various $\alpha$, and dots on them are intersection points between null geodesics and $t=0$, which together form the CISs.
    One can obtain similar plots also for larger $d$'s.}
    \label{fig: surface coincidence}
\end{figure}

In this appendix, the coincidence of the CIS and minimal surface for any ball $A$ is numerically confirmed in AdS black hole geometries.
We treat planar, spherical, and hyperbolic black holes, whose metrics can be rewritten into
\begin{align}
    ds^2 = \frac{1}{z^2}\qty(-f(z) dt^2 + \frac{dz^2}{f(z)}+d \bm x^2),
\end{align}
where the AdS radius is set to 1, and $f(z)$ and $d\bm x^2$ are given as follows, respectively:
\begin{align}
    \mbox{Planar BH:}&\qquad f(z) = 1-\qty(\frac{z}{z_h})^d,\qquad d\bm x^2 = d r^2 + r^2d\Omega_{d-2}^2,\\
    \mbox{Spherical BH:}&\qquad f(z) = 1+z^2-(1+z_h^2)\qty(\frac{z}{z_h})^d,\qquad d\bm x^2 = d r^2 + \sin^2 r d\Omega_{d-2}^2,\\
    \mbox{Hyperbolic BH:}&\qquad f(z) = 1-z^2-(1-z_h^2)\qty(\frac{z}{z_h})^d,\qquad d\bm x^2 = d r^2 + \sinh^2r d\Omega_{d-2}^2.
\end{align}
Since each of the above geometries has its own symmetry, as the boundary ball, we only focus on $A=\{(t,z,\bm x)| r\le \alpha,~t=0\}$.
When $d=2$, all of those geometries are identical to the BTZ black hole, except for whether $r$-direction is compactified or not.
The coincidence $\mathcal{C}_A = \mathcal{E}_A$ for the BTZ geometry is analytically shown in \cite{Hubeny:2012wa}.

First, the minimal surface of $A$, which from the symmetry is written as $z=z(r)$, extremizes the Nambu-Goto action
\begin{align}
    I = \int dr\, \frac{r^{d-2}}{z(r)^{d-1}}\sqrt{1+\frac{z'(r)^2}{f(z(r))}}.
\end{align}
Then, $z(r)$ is determined through
\begin{align}
   z''-\frac{z'^2 \left(2 (d-1) z'+r f'(z)\right)}{2 r f(z)}-z' \left(\frac{(d-2) z'}{z}+\frac{d-1}{r}\right)-\frac{(d-2) f(z)}{z}=0.
\end{align}

On the other hand, $\mathcal C_A$ is computed as a set of intersection points between null geodesics shot from $(t,z,r)=(-\alpha,0,0)$ and time slice $t=0$.
Here the boundary point $(-\alpha,0,0)$ is the bottom vertex of $\Diamond_A$.
Since $\mathcal{C}_A$ is on $\partial\blacklozenge_A$, null geodesics that reach $\mathcal{C}_A$ must not go around the compact directions, $\Omega_{d-2}$.
Thus, we only consider null geodesics whose $\Omega_{d-2}$-coordinates are constant.

The coincidence $\mathcal{C}_A = \mathcal{E}_A$ for various $d$ and $\alpha$ is seen in Fig.\,\ref{fig: surface coincidence}.
This evidence supports that the method in \S\ref{sec: reconstruction} seems to work in reconstructing those black hole spacetimes.

\bibliographystyle{jhep} 
\bibliography{ref}

\providecommand{\href}[2]{#2}\begingroup\raggedright\begin{thebibliography}{10}

\bibitem{Engelhardt:2016wgb}
N.~Engelhardt and G.T.~Horowitz, \emph{{Towards a Reconstruction of General
  Bulk Metrics}},
  \href{https://doi.org/10.1088/1361-6382/34/1/015004}{\emph{Class. Quant.
  Grav.} {\bfseries 34} (2017) 015004}
  [\href{https://arxiv.org/abs/1605.01070}{{\ttfamily 1605.01070}}].

\bibitem{Engelhardt:2016crc}
N.~Engelhardt and G.T.~Horowitz, \emph{{Recovering the spacetime metric from a
  holographic dual}},
  \href{https://doi.org/10.4310/ATMP.2017.v21.n7.a2}{\emph{Adv. Theor. Math.
  Phys.} {\bfseries 21} (2017) 1635}
  [\href{https://arxiv.org/abs/1612.00391}{{\ttfamily 1612.00391}}].

\bibitem{Engelhardt:2016vdk}
N.~Engelhardt, \emph{{Into the Bulk: A Covariant Approach}},
  \href{https://doi.org/10.1103/PhysRevD.95.066005}{\emph{Phys. Rev. D}
  {\bfseries 95} (2017) 066005}
  [\href{https://arxiv.org/abs/1610.08516}{{\ttfamily 1610.08516}}].

\bibitem{Hernandez-Cuenca:2020ppu}
S.~Hern\'andez-Cuenca and G.T.~Horowitz, \emph{{Bulk reconstruction of metrics
  with a compact space asymptotically}},
  \href{https://doi.org/10.1007/JHEP08(2020)108}{\emph{JHEP} {\bfseries 08}
  (2020) 108} [\href{https://arxiv.org/abs/2003.08409}{{\ttfamily
  2003.08409}}].

\bibitem{Burda:2018rpb}
P.~Burda, R.~Gregory and A.~Jain, \emph{{Holographic reconstruction of bubble
  spacetimes}}, \href{https://doi.org/10.1103/PhysRevD.99.026003}{\emph{Phys.
  Rev. D} {\bfseries 99} (2019) 026003}
  [\href{https://arxiv.org/abs/1804.05202}{{\ttfamily 1804.05202}}].

\bibitem{Folkestad:2021kyz}
r.~Folkestad and S.~Hern\'andez-Cuenca, \emph{{Conformal Rigidity from
  Focusing}}, \href{https://doi.org/10.1088/1361-6382/ac27ef}{\emph{Class.
  Quant. Grav.} {\bfseries 38} (2021) 21}
  [\href{https://arxiv.org/abs/2106.09037}{{\ttfamily 2106.09037}}].

\bibitem{Takeda:2021dsl}
D.~Takeda, \emph{{Light-cone cuts and hole-ography: explicit reconstruction of
  bulk metrics}}, \href{https://doi.org/10.1007/JHEP04(2022)124}{\emph{JHEP}
  {\bfseries 04} (2022) 124}
  [\href{https://arxiv.org/abs/2112.11437}{{\ttfamily 2112.11437}}].

\bibitem{Maldacena:2015iua}
J.~Maldacena, D.~Simmons-Duffin and A.~Zhiboedov, \emph{{Looking for a bulk
  point}}, \href{https://doi.org/10.1007/JHEP01(2017)013}{\emph{JHEP}
  {\bfseries 01} (2017) 013}
  [\href{https://arxiv.org/abs/1509.03612}{{\ttfamily 1509.03612}}].

\bibitem{Hubeny:2012wa}
V.E.~Hubeny and M.~Rangamani, \emph{{Causal Holographic Information}},
  \href{https://doi.org/10.1007/JHEP06(2012)114}{\emph{JHEP} {\bfseries 06}
  (2012) 114} [\href{https://arxiv.org/abs/1204.1698}{{\ttfamily 1204.1698}}].

\bibitem{Balasubramanian:2013lsa}
V.~Balasubramanian, B.D.~Chowdhury, B.~Czech, J.~de~Boer and M.P.~Heller,
  \emph{{Bulk curves from boundary data in holography}},
  \href{https://doi.org/10.1103/PhysRevD.89.086004}{\emph{Phys. Rev. D}
  {\bfseries 89} (2014) 086004}
  [\href{https://arxiv.org/abs/1310.4204}{{\ttfamily 1310.4204}}].

\bibitem{Balasubramanian:2014sra}
V.~Balasubramanian, B.D.~Chowdhury, B.~Czech and J.~de~Boer, \emph{{Entwinement
  and the emergence of spacetime}},
  \href{https://doi.org/10.1007/JHEP01(2015)048}{\emph{JHEP} {\bfseries 01}
  (2015) 048} [\href{https://arxiv.org/abs/1406.5859}{{\ttfamily 1406.5859}}].

\bibitem{Headrick:2014eia}
M.~Headrick, R.C.~Myers and J.~Wien, \emph{{Holographic Holes and Differential
  Entropy}}, \href{https://doi.org/10.1007/JHEP10(2014)149}{\emph{JHEP}
  {\bfseries 10} (2014) 149} [\href{https://arxiv.org/abs/1408.4770}{{\ttfamily
  1408.4770}}].

\bibitem{Czech:2014ppa}
B.~Czech and L.~Lamprou, \emph{{Holographic definition of points and
  distances}}, \href{https://doi.org/10.1103/PhysRevD.90.106005}{\emph{Phys.
  Rev. D} {\bfseries 90} (2014) 106005}
  [\href{https://arxiv.org/abs/1409.4473}{{\ttfamily 1409.4473}}].

\bibitem{Roy:2018ehv}
S.R.~Roy and D.~Sarkar, \emph{{Bulk metric reconstruction from boundary
  entanglement}}, \href{https://doi.org/10.1103/PhysRevD.98.066017}{\emph{Phys.
  Rev. D} {\bfseries 98} (2018) 066017}
  [\href{https://arxiv.org/abs/1801.07280}{{\ttfamily 1801.07280}}].

\bibitem{Kabat:2017mun}
D.~Kabat and G.~Lifschytz, \emph{{Local bulk physics from intersecting modular
  Hamiltonians}}, \href{https://doi.org/10.1007/JHEP06(2017)120}{\emph{JHEP}
  {\bfseries 06} (2017) 120}
  [\href{https://arxiv.org/abs/1703.06523}{{\ttfamily 1703.06523}}].

\bibitem{Ryu:2006bv}
S.~Ryu and T.~Takayanagi, \emph{{Holographic derivation of entanglement entropy
  from AdS/CFT}},
  \href{https://doi.org/10.1103/PhysRevLett.96.181602}{\emph{Phys. Rev. Lett.}
  {\bfseries 96} (2006) 181602}
  [\href{https://arxiv.org/abs/hep-th/0603001}{{\ttfamily hep-th/0603001}}].

\bibitem{Ryu:2006ef}
S.~Ryu and T.~Takayanagi, \emph{{Aspects of Holographic Entanglement Entropy}},
  \href{https://doi.org/10.1088/1126-6708/2006/08/045}{\emph{JHEP} {\bfseries
  08} (2006) 045} [\href{https://arxiv.org/abs/hep-th/0605073}{{\ttfamily
  hep-th/0605073}}].

\bibitem{Wall:2012uf}
A.C.~Wall, \emph{{Maximin Surfaces, and the Strong Subadditivity of the
  Covariant Holographic Entanglement Entropy}},
  \href{https://doi.org/10.1088/0264-9381/31/22/225007}{\emph{Class. Quant.
  Grav.} {\bfseries 31} (2014) 225007}
  [\href{https://arxiv.org/abs/1211.3494}{{\ttfamily 1211.3494}}].

\bibitem{Headrick:2014cta}
M.~Headrick, V.E.~Hubeny, A.~Lawrence and M.~Rangamani, \emph{{Causality \&
  holographic entanglement entropy}},
  \href{https://doi.org/10.1007/JHEP12(2014)162}{\emph{JHEP} {\bfseries 12}
  (2014) 162} [\href{https://arxiv.org/abs/1408.6300}{{\ttfamily 1408.6300}}].

\bibitem{Cardy:2016fqc}
J.~Cardy and E.~Tonni, \emph{{Entanglement hamiltonians in two-dimensional
  conformal field theory}},
  \href{https://doi.org/10.1088/1742-5468/2016/12/123103}{\emph{J. Stat. Mech.}
  {\bfseries 1612} (2016) 123103}
  [\href{https://arxiv.org/abs/1608.01283}{{\ttfamily 1608.01283}}].

\bibitem{Kelly:2013aja}
W.R.~Kelly and A.C.~Wall, \emph{{Coarse-grained entropy and causal holographic
  information in AdS/CFT}},
  \href{https://doi.org/10.1007/JHEP03(2014)118}{\emph{JHEP} {\bfseries 03}
  (2014) 118} [\href{https://arxiv.org/abs/1309.3610}{{\ttfamily 1309.3610}}].

\bibitem{Czech:2015qta}
B.~Czech, L.~Lamprou, S.~McCandlish and J.~Sully, \emph{{Integral Geometry and
  Holography}}, \href{https://doi.org/10.1007/JHEP10(2015)175}{\emph{JHEP}
  {\bfseries 10} (2015) 175}
  [\href{https://arxiv.org/abs/1505.05515}{{\ttfamily 1505.05515}}].

\bibitem{Czech:2015kbp}
B.~Czech, L.~Lamprou, S.~McCandlish and J.~Sully, \emph{{Tensor Networks from
  Kinematic Space}}, \href{https://doi.org/10.1007/JHEP07(2016)100}{\emph{JHEP}
  {\bfseries 07} (2016) 100}
  [\href{https://arxiv.org/abs/1512.01548}{{\ttfamily 1512.01548}}].

\bibitem{Czech:2019hdd}
B.~Czech, Y.D.~Olivas and Z.-z.~Wang, \emph{{Holographic integral geometry with
  time dependence}}, \href{https://doi.org/10.1007/JHEP12(2020)063}{\emph{JHEP}
  {\bfseries 12} (2020) 063}
  [\href{https://arxiv.org/abs/1905.07413}{{\ttfamily 1905.07413}}].

\bibitem{Casini:2003ix}
H.~Casini, \emph{{Geometric entropy, area, and strong subadditivity}},
  \href{https://doi.org/10.1088/0264-9381/21/9/011}{\emph{Class. Quant. Grav.}
  {\bfseries 21} (2004) 2351}
  [\href{https://arxiv.org/abs/hep-th/0312238}{{\ttfamily hep-th/0312238}}].

\bibitem{Hubeny:2007xt}
V.E.~Hubeny, M.~Rangamani and T.~Takayanagi, \emph{{A Covariant holographic
  entanglement entropy proposal}},
  \href{https://doi.org/10.1088/1126-6708/2007/07/062}{\emph{JHEP} {\bfseries
  07} (2007) 062} [\href{https://arxiv.org/abs/0705.0016}{{\ttfamily
  0705.0016}}].

\bibitem{Hubeny:2012ry}
V.E.~Hubeny, \emph{{Extremal surfaces as bulk probes in AdS/CFT}},
  \href{https://doi.org/10.1007/JHEP07(2012)093}{\emph{JHEP} {\bfseries 07}
  (2012) 093} [\href{https://arxiv.org/abs/1203.1044}{{\ttfamily 1203.1044}}].

\bibitem{Engelhardt:2013tra}
N.~Engelhardt and A.C.~Wall, \emph{{Extremal Surface Barriers}},
  \href{https://doi.org/10.1007/JHEP03(2014)068}{\emph{JHEP} {\bfseries 03}
  (2014) 068} [\href{https://arxiv.org/abs/1312.3699}{{\ttfamily 1312.3699}}].

\bibitem{Engelhardt:2015dta}
N.~Engelhardt and S.~Fischetti, \emph{{Covariant Constraints on the
  hole-ography}},
  \href{https://doi.org/10.1088/0264-9381/32/19/195021}{\emph{Class. Quant.
  Grav.} {\bfseries 32} (2015) 195021}
  [\href{https://arxiv.org/abs/1507.00354}{{\ttfamily 1507.00354}}].

\bibitem{Lin:2016fqk}
J.~Lin, \emph{{A Toy Model of Entwinement}},
  \href{https://arxiv.org/abs/1608.02040}{{\ttfamily 1608.02040}}.

\bibitem{Balasubramanian:2016xho}
V.~Balasubramanian, A.~Bernamonti, B.~Craps, T.~De~Jonckheere and F.~Galli,
  \emph{{Entwinement in discretely gauged theories}},
  \href{https://doi.org/10.1007/JHEP12(2016)094}{\emph{JHEP} {\bfseries 12}
  (2016) 094} [\href{https://arxiv.org/abs/1609.03991}{{\ttfamily
  1609.03991}}].

\bibitem{Erdmenger:2019lzr}
J.~Erdmenger and M.~Gerbershagen, \emph{{Entwinement as a possible alternative
  to complexity}}, \href{https://doi.org/10.1007/JHEP03(2020)082}{\emph{JHEP}
  {\bfseries 03} (2020) 082}
  [\href{https://arxiv.org/abs/1910.05352}{{\ttfamily 1910.05352}}].

\bibitem{Craps:2022pke}
B.~Craps, M.~De~Clerck and A.~Vilar~L\'opez, \emph{{Definitions of
  entwinement}},  \href{https://arxiv.org/abs/2211.17253}{{\ttfamily
  2211.17253}}.

\bibitem{Chen_2014}
Y.~Chen and G.~Vidal, \emph{Entanglement contour}, .

\bibitem{Wen:2018whg}
Q.~Wen, \emph{{Fine structure in holographic entanglement and entanglement
  contour}}, \href{https://doi.org/10.1103/PhysRevD.98.106004}{\emph{Phys. Rev.
  D} {\bfseries 98} (2018) 106004}
  [\href{https://arxiv.org/abs/1803.05552}{{\ttfamily 1803.05552}}].

\bibitem{Wen:2019iyq}
Q.~Wen, \emph{{Formulas for Partial Entanglement Entropy}},
  \href{https://doi.org/10.1103/PhysRevResearch.2.023170}{\emph{Phys. Rev.
  Res.} {\bfseries 2} (2020) 023170}
  [\href{https://arxiv.org/abs/1910.10978}{{\ttfamily 1910.10978}}].

\bibitem{Rolph:2021nan}
A.~Rolph, \emph{{Local measures of entanglement in black holes and CFTs}},
  \href{https://doi.org/10.21468/SciPostPhys.12.3.079}{\emph{SciPost Phys.}
  {\bfseries 12} (2022) 079}
  [\href{https://arxiv.org/abs/2107.11385}{{\ttfamily 2107.11385}}].

\end{thebibliography}\endgroup
\end{document}